\documentclass[aps,pre]{revtex4}

 \usepackage{amsmath,amssymb,amsthm,mathrsfs,amsfonts,dsfont} 
\usepackage{dcolumn}
\usepackage{graphicx}
\usepackage{amsbsy}
\usepackage{bm}
\usepackage{algorithm}
\usepackage{algorithmicx}
\usepackage{algpseudocode}
\usepackage{physics}
\usepackage{xcolor}

\newcommand{\blue}[1]{{\color{black}#1}}

\newcommand{\lrbka}[1]{\left\langle #1 \right\rangle}

\def\X{\mathbf{X}}
\def\bxi{\boldsymbol{\xi}}
\def\I{\mathbf{I}_d}
\def\bmu{\boldsymbol{\mu}}
\def\N{\mathcal{N}}
\def\A{\mathcal{A}}

\def\T{\top}
\def\J{\mathbf{J}}
\def\S{\mathbb{S}}

\begin{document}

\title{Nonequilbrium physics of generative diffusion models}

\author{Zhendong Yu$^{1}$}
\author{Haiping Huang$^{1,2}$}
\email{huanghp7@mail.sysu.edu.cn}
\affiliation{$^{1}$PMI Lab, School of Physics,
Sun Yat-sen University, Guangzhou 510275, People's Republic of China}
\affiliation{$^{2}$Guangdong Provincial Key Laboratory of Magnetoelectric Physics and Devices, Sun Yat-sen University, Guangzhou 510275, People’s Republic of China}
\date{\today}

\begin{abstract}
Generative diffusion models apply the concept of Langevin dynamics in physics to machine leaning, attracting a lot of interests from engineering, statistics and physics, but a complete picture about inherent mechanisms is still lacking. In this paper, we provide a transparent physics analysis of diffusion models, formulating the fluctuation theorem, entropy production, equilibrium measure, and Franz-Parisi potential to understand the dynamic process and intrinsic phase transitions. Our analysis is rooted in a path integral representation of both forward and backward dynamics, and in treating the reverse diffusion generative process as a statistical inference, where the time-dependent state variables serve as quenched disorder akin to that in spin glass theory. Our study thus links stochastic thermodynamics, statistical inference and geometry based analysis together to yield a coherent picture about how the generative diffusion models work.
\end{abstract}

\maketitle

\section{Introduction}
Neural network based machine learning has triggered a lot of research interests in a variety of fields~\cite{DL-2016,Huang-2022}. One of current active directions is the generative diffusion models (GDMs)~\cite{NEDM-2015,DDPM-2020,Song-2019,Song-2020}, which are rooted in nonequilibrium physics~\cite{Risken-1996,Van-2007}. Forward and backward stochastic differential equations (SDEs, or Langevin equations in physics) are used; the forward part is to diffuse a data sample (e.g., a real image) into a Gaussian white noise distribution, after that, taking a sample from this Gaussian white noise distribution starts the backward process, driven by the gradient of log-state-likelihood, and finally this reverse Langevin dynamics collapses onto a real sample subject to the true data distribution, thereby completing the unsupervised data generation. This process is in essence a physical process, whose cornerstone is nonequilibrium dynamics,  a central topic of statistical physics~\cite{Risken-1996,Van-2007,Seifert-2012,Peliti-2021,Zou-2024}.

Recent interests from physics community focused on symmetry breaking in the diffusion process~\cite{Mezard-2023,SSB-DM-2023,Mezard-2024}, Bayes-optimal denoising interpretation of the  GDMs~\cite{Lenka-2023}, reformulation as equilibrium statistical mechanics~\cite{Free-2023}, and path integral representation of the stochastic trajectories~\cite{Path-2024}.
We remark that the symmetry breaking concept in unsupervised learning (GDM is one type of unsupervised learning) has been introduced and analyzed in earlier works~\cite{Huang-2019,Huang-2020}. In this work, we define a high-dimensional Gaussian mixture data model that allows for analytic studies. Although recent works also studied this Gaussian mixture data~\cite{Mezard-2024}, our main contributions in this work are distinct from recent works in the following three aspects. Our formulas apply to the high-dimensional dynamics, but we show one-dimensional examples for demonstration of the concept.

First, the GDM is analyzed through the concept of entropy production, especially for the reverse generative process. The derived formulas apply to the high-dimensional case. For a simple demonstration, the environmental and system entropy changes are explicitly computed, obeying the fluctuation theorem. The ensemble entropy production rate is also calculated for an example of one-dimensional diffusion, displaying distinct qualitative behaviors in both forward and backward processes. We also demonstrate the same dynamic state probability for both processes, and the probability currents have the same magnitude but opposite directions for the forward and reverse processes.

Second, we derive a generalized form of statistical inference for our current data model, where a temperature-like parameter is set by a variance matrix. Moreover, we prove that for this generalized form, the free energy is exactly the potential energy derived previously in Ref.~\cite{SSB-DM-2023}. By an intuitive exploration of the qualitative shape of the potential (or equivalently the free energy of the statistical inference problem), we get the speciation transition point derived previously by an alternative method of Landau expansion or overlap dynamics~\cite{Mezard-2024}. \blue{Moreover, we analyze the case of previously-unexplored one-dimensional diffusion example of arbitrary data mean and variance, and reveal a phase diagram of symmetry breaking, symmetry un-breaking and unstable symmetry breaking phases}. Hence, we unify relevant results in our generalized form of statistical inference.

Third, we provide a geometry oriented way to look at a recently discovered collapse transition in the reverse diffusion process. In previous works, an empirical distribution for the reverse dynamic state
is assumed, and the collapse transition is thus related to the number of data points used to represent the empirical distribution and the dimensionality of the dynamic state as well. In contrast, we start from the statistical inference standpoint without using the empirical distribution, and define a geometry measure, i.e., Franz-Parisi potential (see the first application in neural networks~\cite{Huang-2014}) to capture the potential emergence of hidden structures in the conditional probability of the statistical inference. We provide an effective description in a one-dimensional example. The complicated yet tractable computation of the high dimensional case is left for future works. Our work focuses on a pure understanding of an analytically tractable GDM, without taking into account the algorithmic design, which typically requires training a complex neural network for non-Gaussian real dataset.

The paper is organized as follows. We first introduce the forward diffusion process, together with the fluctuation theorem derived from the forward process, and concepts of stochastic entropy and ensemble average. Then we derive the reverse generative dynamics applied to generate the real data samples in machine learning, and introduce in details the concept of potential energy and free energy to analyze the denoising process formulated as a statistical inference. We also derive the fluctuation theorem and entropy production rate for the reverse process, together with Franz-Parisi potential applied to analyze how fragmented the configuration space for the inference is as the time approaches the starting point of the forward dynamics. We finally summarize our studies and make future perspectives in the last section.

\section{Nonequilibrium physics of forward diffusion process}
\subsection{Forward diffusion dynamics}
A classic example of stochastic dynamics is the well-known Brownian motion, whose dynamics is called the Langevin dynamics. Consistence between Brownian motion and thermodynamics has been established in 1905~\cite{Albert-1905}.  Current AI studies make nonequilibrium physics of Langevin dynamics regain intense research interests~\cite{Zou-2024,Path-2024}.
In the forward process, we use Ornstein-Ulhenbeck (OU) process~\cite{Risken-1996} to turn a real data point into a white noise, which is detailed as a high dimensional SDE:
\begin{equation}\label{sde0}
\dot{\X}=-\X+\sqrt{2}\bxi,
\end{equation}
where $\X\in\mathbb{R}^d$ and $\bxi\in\mathbb{R}^d$ are time-dependent high dimensional state and noise quantities, respectively, and $\bxi$ is a Gaussian white noise with correlation $\lrbka{\xi_i(t)\xi_j(t')}=\delta_{ij}\delta(t-t')$. Given the initial condition $\X(0)\equiv\X_0$, the above SDE has a solution:
\begin{equation}
\X_t=e^{-t}\X_0+\sqrt{2}e^{-t}\int_0^{t}dse^{s}\bxi(s),
\end{equation}
from which, $\X(t)\equiv\X_t$ is clearly a Gaussian random variable, which can be reformulated as the following form using independent standard Gaussian random variable:
\begin{equation}\label{evo1}
\X_t = e^{-t} \X_0 + \sqrt{1-e^{-2t}} \mathbf{Z}_t,
\end{equation}
where $\mathbf{Z}_t$ is the standard Gaussian random variable, and the variance of $\X_t$ is given by $1-e^{-2t}$. In the following, we make no difference between $\X(t)$ and $\X_t$.

\begin{figure}
\centering
\includegraphics[width=0.9\textwidth]{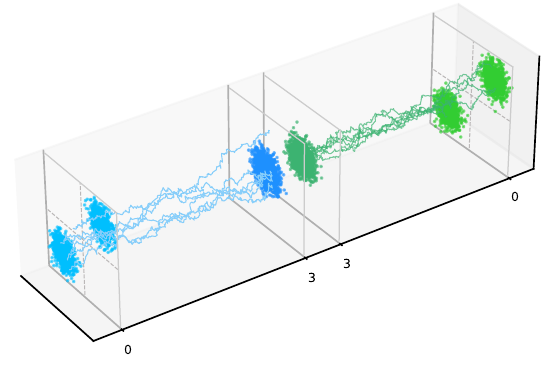}
\caption{A schematic illustration of the generative diffusion process of two-dimensional Gaussian mixture data. The forward process from time $t = 0$ to $t = 3$ is shown together with the reverse process from $t = 3$ back to $t = 0$. The gradient of log-state-likelihood can be analytically estimated as the state probability is given by $p({X_t},t) = \frac{1}{2}\N\left( {\bmu{e^{ - t}},\I} \right) + \frac{1}{2}\N\left( { - \bmu{e^{ - t}},\I} \right)$ .}
\label{fig1}
\end{figure}

For simplicity, we choose the distribution of the data as a Gaussian mixture of two classes (e.g., two kinds of images):
\begin{equation}
   p(\X_0) = \frac{1}{2}\mathcal{N}\left( \bmu,\I \right) + \frac{1}{2}\mathcal{N}\left( -\bmu,\I \right) ,
\end{equation}
 where $\bmu$ is the $d$-dimensional constant vector, and $\I$ is the $d$-dimensional identity covariance matrix. In most parts of this paper, we consider this simple Gaussian mixture with unit variance. The more general case of non-unit variance is also discussed when necessary. It is also straightforward to generalize our analysis to multiple classes. 
Then at time $t$, the probability distribution $p(\X_t,t)$ can be calculated by a convolution~\cite{GMDM-2023} 
 \begin{equation}\label{eq5}
 \begin{aligned}
 p(\X_t,t)&=\int d\X_0p(\X_0)p(\X_t|\X_0)\\
 &=\int d\X_0p(\X_0)\mathcal{N}(\X_t;\X_0e^{-t},\Sigma_t\I)\\
 &=\frac{1}{2}\mathcal{N}(\X_t;\bmu_t,\I)+\frac{1}{2}\mathcal{N}(\X_t;-\bmu_t,\I),
 \end{aligned}
 \end{equation}
where $\bmu_t\equiv\bmu e^{-t}$, and $\Sigma_t\equiv1-e^{-2t}$. Representative trajectories of the forward process are shown in Fig.~\ref{fig1}.

\subsection{Fluctuation theorem for the forward diffusion}
Because of stochasticity, the trajectories
are not differentiable any more in general. A specific time-discretization scheme for the stochastic differential equation must be carefully chosen, for which the stochastic calculus 
is established (see details below).
Then we would derive the path probability of $\{\X_t\}$ given an initial point $\X_0$. To do this, we have to specify the discretization scheme 
for the above SDEs. We first define random variable $\mathbf{W}(t)$ for the Wiener process as follows:
\begin{equation}
\mathbf{W}(t)=\int_{t_0}^{t}dt'\bxi(t')=\int_{t_0}^{t}d\mathbf{W}.
\end{equation}
The stochastic integral can be expressed as $\int_{t_0}^{t}dt'\bxi(t')f(\X_t,t)=\int_{t_0}^{t}d\mathbf{W}f(\X_t,t)$, called the Riemann–Stieltjes integral as well. In studies of SDEs, we have two commonly used
conventions to represent this stochastic integral.
The first one is the Ito convention, or the initial point scheme. More precisely, the Riemann–Stieltjes integral is calculated as
\begin{equation}
\int_{t_0}^{t}d\mathbf{W}f(\X_t,t)=\lim_{dt\to0}\sum_{k=0}^N\left[\mathbf{W}(t_k+dt)-\mathbf{W}(t_k)\right]f(\mathbf{X}(t_k),t_k),
\end{equation}
where $t_k=t_0+kdt$, and $N=\frac{t-t_0}{dt}$. In general, we interpolate the time between $t_k$ and $t_{k+1}$ as $\tau=(1-\lambda)t_k+\lambda t_{k+1}$. Therefore, the Ito convention corresponds to $\lambda=0$. This stochastic integral is clearly dependent on the discretization scheme~\cite{Peliti-2021}.

The second one is the Stratonovich convention, i.e., mid-point scheme with $\lambda=1/2$. Then, we have the following expression:
\begin{equation}
\int_{t_0}^{t}d\mathbf{W}f(\X_t,t)=\lim_{dt\to0}\sum_{k=0}^N\left[\mathbf{W}(t_k+dt)-\mathbf{W}(t_k)\right]f\left([\X(t_k+dt)+\X(t_k)]/2,t_k+dt/2\right).
\end{equation}

Next we rewrite Eq.~\eqref{sde0} as follows,
\begin{equation}
d\X_t=f(\X_t,t)dt+\sqrt{2}d\mathbf{W},
\end{equation}
where $dt$ is a small step size as used before. According to the general discretization, we have
\begin{equation}\label{Meq}
\frac{{{\X_{t + dt}} - {\X_t}}}{{dt}} = f\left((1 - \lambda ){\X_t} + \lambda {\X_{t + dt}},t + \lambda dt\right) + \boldsymbol{\eta}_t,
\end{equation}
where $\eta_{i,t}\sim\mathcal{N}(0,2/dt)$. In the following we can take $t+dt$ as $t'$, or $t'-t=dt$ in the Markovian process defined by Eq.~\eqref{Meq}; $dt$ is the unit of the above discretization.

We next use the following probability transformation identity:
\begin{equation}
   P({\X_{t'}},t'|{\X_t},t) = P(\boldsymbol{\eta}_t )\left| {\frac{{\partial \boldsymbol{\eta}_t }}{\partial \X_{t'}}} \right|,
\end{equation}
where the determinant $\left| {\frac{{\partial \boldsymbol{\eta}_t }}{\partial \X_{t'}}} \right|$ is a Jacobian measuring the change of volume for the transformed
 probability density. This Jacobian can be easily computed using Eq.~\eqref{Meq}.
 \begin{equation}
\begin{aligned}
& \left| {\frac{{\partial\boldsymbol{\eta}_t }}{{\partial {\X_{t'}}}}} \right| = \left| {\frac{{\partial \left( {\frac{{{\X_{t + dt}} - {\X_t}}}{{dt}} - f\left( {(1 - \lambda ){\X_t} + \lambda {\X_{t + dt}},t + \lambda dt} \right)} \right)}}{{\partial {\X_{t + dt}}}}} \right| \\
&  \propto {e^{ - \lambda \nabla  \cdot fdt}},\label{div}
\end{aligned}
\end{equation}
where $\I$ is an $d$-dimensional identity matrix,  we have used the Taylor expansion ($dt\to0$) and the matrix identity $\operatorname{det}\left(e^{\mathbf{K}}\right)=e^{\operatorname{Tr} \mathbf{K}}$. Finally, based on the known form of the white noise distribution, we have the following infinitesimal propagator:
\begin{equation}
    P({\X_{t+dt},t+dt}|{\X_t},t)  \propto e^{ - \lambda \nabla  \cdot fdt}e^{ - \frac{{{{\left| {{{\dot \X}_t} - f({\X_t},t)} \right|}^2}}}{4}dt}.
\end{equation}
Using the Markovian chain property, we get the conditional trajectory probability:
\begin{equation}
\begin{aligned}
    P({X([T])}|{\X_0}) &= \prod\limits_{t'} {P({\X_{t'},t'}|{\X_t},t)} \\
    &   = \prod\limits_{t'} e^{ - \lambda \nabla  \cdot fdt}e^{ - \frac{{{{\left| {{{\dot \X}_t} - f({\X_t},t)} \right|}^2}}}{4}dt}   \\
    &   \propto {e^{\int_0^T {[ - \lambda \nabla  \cdot f - \frac{{{{\left| {{{\dot \X}_t} - f({\X_t},t)} \right|}^2}}}{4}]\mathop  \odot \limits^\lambda  dt} }},
    \end{aligned}
 \end{equation}
 where $X([T])\equiv\{\X_T,\ldots,\X_1\}$ specifying a trajectory starting from $\X_0$, $T$ denotes the length of the individual trajectory, and $\mathop  \odot \limits^\lambda$ denotes the corresponding $\lambda$-convention for the stochastic integral.
An alternative way to get the same result is to use the following property of Dirac delta function:
\begin{equation}
\delta\left(\X(t)-\X_\xi(t)\right)=\delta\left(\mathbf{O}[\X(t)]\right)\operatorname{det}\frac{\delta \mathbf{O}}{\delta \X},
\end{equation}
where $\X_\xi(t)$ is a solution of $\mathbf{O}[\X]=\dot{\X}-f-\bxi=0$. Therefore, $P(X[T]|\X_0)=\lrbka{\prod_{t}\delta(\X(t)-\X_\xi(t))}_{\{\bxi_t\}}$.

Now, taking $f=-\X$ [see Eq.~\eqref{sde0}], we can derive the trajectory probability for the forward OU process
\begin{equation}
\begin{aligned}
    P\left( {X([T] )|{\X_0}} \right) &= \prod\limits_{t'} {P\left( {{\X_{t'}},t'\vert \X_t,t} \right)} \\
    & \propto \exp \left( - \int\limits_0^T \left[\frac{1}{4}{{\left( {\dot \X + \X} \right)}^2}  - \lambda d\right]\mathop  \odot \limits^\lambda  dt \right),\label{mav}
\end{aligned}
\end{equation}
where $d$ is the dimensionality of the dynamics, and the term in the exponent is called the action in physics for the path probability, and the integrand inside the time integral of the action is called the Lagrangian $\mathcal{L}$~\cite{JPA-2017}, and the optimal path of maximal trajectory probability is determined by the Euler-Lagrange equation $\frac{d}{dt}\left(\frac{\partial\mathcal{L}}{\partial\dot{\X}}\right)-\frac{\partial\mathcal{L}}{\partial\X}=0$. 

Next, we consider a reverse dynamics, i.e., $\tilde{\X}(s)=\X(t)$, where $s=T-t$, and $T$ is the time length of the trajectory. It is clear that $\tilde{\X}(0)=\X(T)$, and $\tilde{\X}(T)=\X(0)$. We have then $\dot{\tilde{\X}}(s)=-\dot{\X}(t)$. In analogy to the forward trajectory, the path probability of the backward trajectory given the initial point is given by
\begin{equation}
    P\left[ {\tilde X([T] )\mid \tilde \X_0} \right] = \N\exp [ - \A(\tilde X([T] ))],\\
    \end{equation}
where $\N$ is a normalization factor, $\tilde \X_t\equiv\tilde \X(t)$ (similar for $\X_t$), and the action reads,
\begin{equation}
\begin{aligned}
  \A(\tilde X([T] ) )&= \int_0^T d s\mathop  \odot \limits^\lambda  \left[ {{{(\dot {\tilde \X} + \tilde \X)}^2}/4 - \lambda d} \right] \\
  &= \int_0^T d t \mathop  \odot \limits^{1 - \lambda } \left[ {{{( - \dot \X + \X)}^2}/4 - \lambda d} \right].
  \end{aligned}
\end{equation}
Note that the time reversal changes the $\lambda$-convention to $(1-\lambda)$-convention~\cite{JPA-2017}. The ratio between the conditional path probabilities is thus given by
\begin{equation}\label{fsm1}
\begin{aligned}
  \ln \left[\frac{{P\left[ {X([T])\mid \X_0} \right]}}{{P\left[ {\tilde X([T] )\mid \tilde \X_0} \right]}}\right] &= \A(\tilde X([T] )) - \A(X([T] ))\\
  &   =  - \int_0^T d t\mathop  \odot \limits^{1 - \lambda } \left[\frac{{  \dot{\X}\cdot\X}}{2}\right] - \int_0^T d t\mathop  \odot \limits^\lambda  \left[\frac{{\dot {\X}\cdot\X}}{2}\right]\\
  & =   \int_0^T d t\mathop  \odot \limits^{\frac{1}{2}} [ - \dot \X\cdot \X],\\
  \end{aligned}
\end{equation}
where the action for the direct dynamics can be read off from Eq.~\eqref{mav}, the calculation in the last step leads to the result independent of $\lambda$ (or equivalently $\lambda=\frac{1}{2}$). This result can be interpreted as heat dissipated into the environment, since in an overdamped system, the product of total mechanical force and displacement equals to dissipation~\cite{Seifert-2012}. Then we identify the following entropy change of environment:
\begin{equation} \label{smf}
  \Delta {S_E} = \int_0^T d t\mathop  \odot \limits^{\frac{1}{2}} [ - \dot \X\cdot\X],
\end{equation}
where we have assumed the temperature for the forward OU process equals to one.

In addition, the starting and final states in the forward diffusion process can be treated as equilibrium states, subject to an analytic form of distribution (in fact they are Gaussian mixture). Then if we define a stochastic or trajectory-dependent entropy of the system as $S(t)=-\ln p(\X_t,t)$, we can derive the entropy change of the system as follows,
\begin{equation}\label{forward}
\begin{aligned}
  \Delta S &= \ln \left[\frac{{p\left( {\X(0)},0 \right)}}{{p(\X(T),T)}}\right]\\
  &  = \ln \left[ {\frac{{\exp ( - \frac{{{{(\X(0) - \bmu)}^2}}}{2}) + \exp ( - \frac{{{{(\X(0) + \bmu)}^2}}}{2})}}{{\exp ( - \frac{{{{(\X(T) - \bmu{e^{ - T}})}^2}}}{2}) + \exp ( - \frac{{{{(\X(T) + \bmu {e^{ - T}})}^2}}}{2})}}} \right].  
  \end{aligned}
\end{equation}
We obtain then the total entropy change:
\begin{equation}\label{stf}
\begin{aligned}
    \Delta {S_{tot}} &= \Delta {S_E} + \Delta S\\
    &= \ln (\frac{{P\left[ {X([T] )} \right]}}{{P\left[ {\tilde X([T] )} \right]}}),
    \end{aligned}
\end{equation}
which implies that the ratio of the path probabilities is exactly $e^{\Delta S_{tot}}$. Note that in Eq.~\eqref{stf} $X([T])$ includes the initial state $X(0)$. Equation~\eqref{stf} is the well-known detailed fluctuation theorem, while the integral fluctuation theorem can be readily derived as
\begin{equation}\label{equ}
\begin{aligned}
    \left\langle {{e^{ - \Delta {S_{tot}}}}} \right\rangle  &= \int {P\left[ {X([T] )} \right]{e^{ - \Delta {S_{tot}}}}dX([T])} \\
    & = \int {P\left[ {\tilde X([T] )} \right]d\tilde X([T] )} \\
    & = 1.
\end{aligned}
\end{equation}
From the integral fluctuation theorem, one can derive the stochastic second thermodynamics law $\lrbka{\Delta S_{tot}}\geq0$ based on the convexity of the exponential function.

\begin{figure}
\centering
\includegraphics[width=0.7\textwidth]{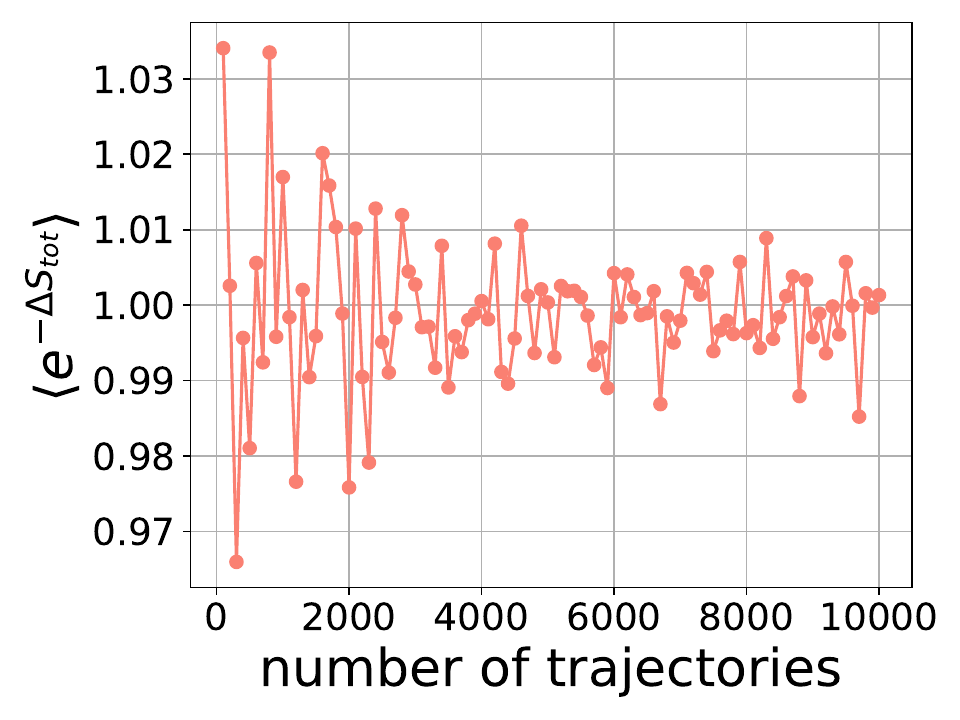}
\caption{Ensemble average of $e^{ - \Delta {S_{tot}}}$ as a function of the number of trajectories used to do the average, simulated by solving the forward OU process. $\mu = 1$, and other parameters are detailed in Appendix~\ref{app-a}.}
\label{forwardS}
\end{figure}

Finally we verify the integral fluctuation theorem in the forward diffusion models (see Fig.~\ref{forwardS}) in an example of one-dimensional OU process. As the number of trajectories gets
large, the trajectory average $\lrbka{\Delta S_{tot}}$ converges to one, as predicted by the theory. Each contribution of the entropy change is shown in Fig.~\ref{forward}. Some trajectories bear a negative entropy change, i.e., entropy decreases during the evolution, but on average, the stochastic second law of thermodynamics is still valid. Experimental details to get the entropy contribution are given in Appendix~\ref{app-a}.

\begin{figure}
\centering
\includegraphics[width=0.85\textwidth]{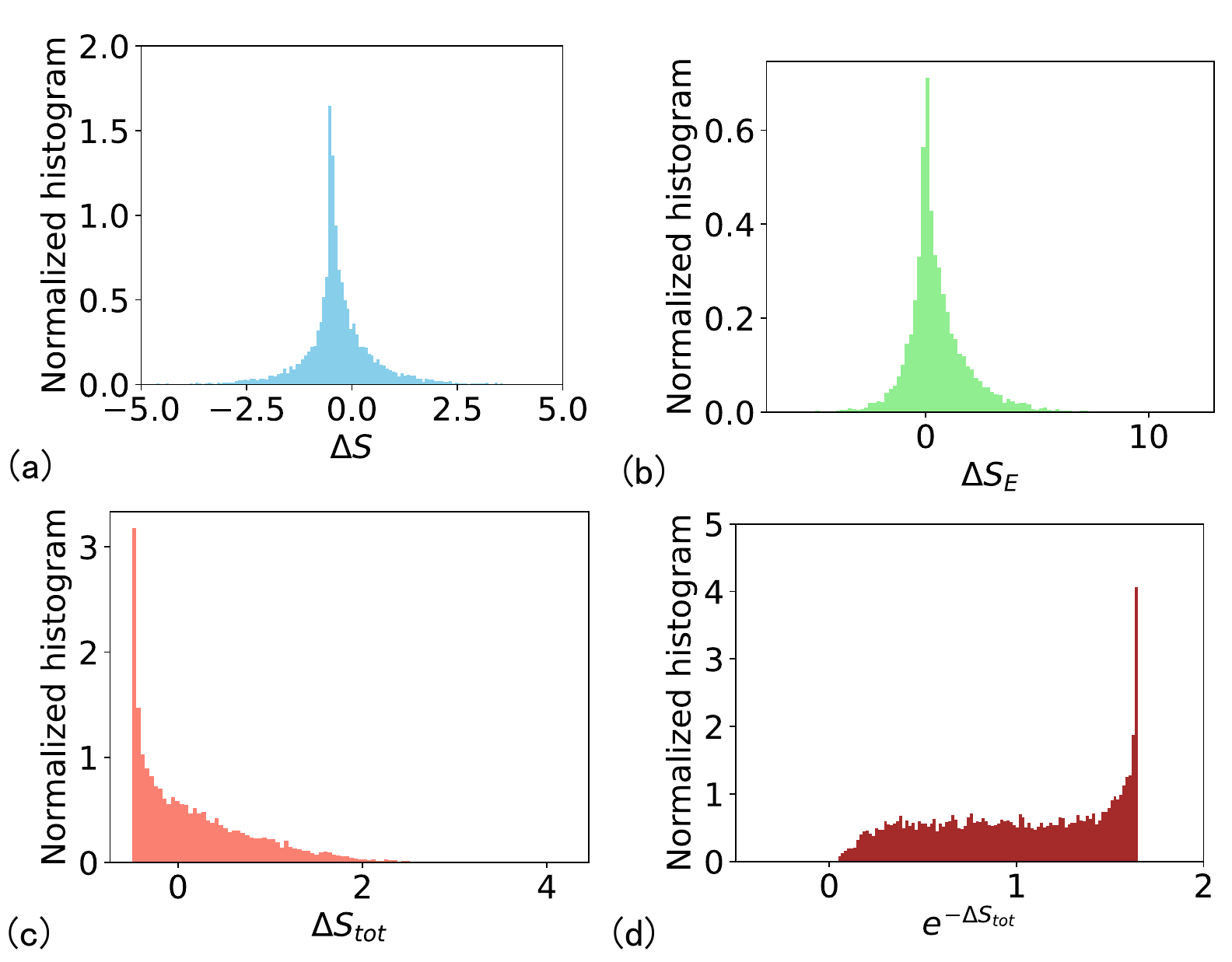}
\caption{ Normalized histogram of entropy changes estimated from $10\,000$ trajectories for the forward OU process. $\mu = 1$, and other parameters are detailed in Appendix~\ref{app-a}. (a) Statistics of system entropy change. (b) Statistics of environment entropy change. (c) Statistics of the total entropy change. (d) Statistics of $e^{-\Delta S_{\text {tot }}}$. }
\label{forward}
\end{figure}

\subsection{Rate of stochastic entropy production}\label{sent}
We first recall the Fokker-Planck equation (FPE) corresponding to the OU forward process under the Ito convention as follows,
\begin{equation}\label{flow}
\begin{aligned}
 \frac{{\partial p(\X_t,t)}}{{\partial t}} &=  - \nabla  \cdot \left[f({\X_t},t)p(\X_t,t)\right] + \sum\limits_{i = 1}^d \frac{{{\partial ^2}p(\X_t,t)}}{{\partial {X_i}\partial {X_i}}}\\
 & =  - \nabla  \cdot \J,
\end{aligned}
\end{equation}
where the probability current reads $\J = f({\X_t},t)p({\X_t},t) - \nabla p({\X_t},t)$, and we have written the force term in Eq.~\eqref{sde0} as a general function $f(\X_t,t)$ which we will specify in the reverse generative dynamics as well. The FPE is an equation of probability conservation~\cite{Risken-1996}.

We then define the stochastic entropy as before~\cite{Seifert-2005} 
\begin{equation}\label{STP}
    \S(t) =  - \ln p(\X_t,t),
\end{equation}
where we set $k_B=1$ in our paper. The rate of the stochastic entropy can be derived directly:
\begin{equation}\label{rS}
\begin{aligned}
   \dot \S(t ) &=  - \frac{{{\partial _t }p(\X_t,t )}}{{p(\X_t,t )}} - \frac{{\nabla p(\X_t,t )}}{{p(\X_t,t)}}\dot \X\\
   &   =  - \frac{{{\partial _t }p(\X_t,t )}}{{p(\X_t,t )}} - \frac{{f(\X_t,t)p(\X_t,t) - \J}}{{p(\X_t,t )}}\dot \X\\
   &  =  - \frac{{{\partial _t }p(\X_t,t )}}{{p(\X_t,t)}} - f(\X_t,t)\cdot\dot \X + \frac{\J\cdot\dot \X}{p(\X_t,t)}.
\end{aligned}
\end{equation}
To derive the above equation, we have used the expression of the probability current. The second term in the last equality of Eq.~\eqref{rS} is actually the rate of heat dissipation to the environment, i.e., $\dot q(t)=f(\X_t,t)\dot \X=\dot{\S}_E(t)$ (notice that we have set the unit temperature). Then, we can define the total entropy production rate $\dot \S_{\rm{tot}}(t ) = {\dot \S}_{E}(t ) + \dot \S(t )$, and as a result, $\dot \S_{\rm{tot}}(t )$ reads,
\begin{equation}
    {{\dot \S}_{{\rm{tot}}}}(t) =   - \frac{{{\partial _t }p(\X_t,t )}}{{p(\X_t,t )}} + \frac{\J\cdot\dot \X}{p\X_t,t )}.
\end{equation}

\subsection{Ensemble entropy production rate}\label{eep}
We first define the ensemble average of the stochastic entropy as follows~\cite{Tome-2006}:
\begin{equation}\label{SEP}
S(t) \equiv\lrbka{\S(t)}= - \int {p({\X_t},t)\ln p({\X_t},t)d\X_t}.
\end{equation}
The rate of entropy change of the system can be readily expanded by inserting the FPE as follows,
\begin{equation}
\begin{aligned}
& \frac{d S(t)}{d t}=\frac{d}{d t}\left(-\int p\left(\X_t, t\right) \ln p\left(\X_t, t\right) d \X_t\right) \\
    &=  - \int {\sum\limits_{i = 1}^d {\frac{{\partial \ln p({\X_t},t)}}{{\partial {X_i}}}} {J_i}} d\X_{t},
\end{aligned}
\end{equation}
where ${{J_i}}$ is the $i$-th component of $\J$, and we have used the fact that $\int d\X_t\frac{\partial p\left(\X_t, t\right)}{\partial t}=0$, or at the boundary, the current vanishes. According to the definition of the probability current $\J$ [see its expression below Eq.~\eqref{flow}], we get its component $J_i=f_ip-p\frac{\partial\ln p}{\partial X_i}$, and replace $\frac{\partial\ln p}{\partial X_i}$ by $f_i-J_i/p$, where $f_i$ is the $i$-th component of the high dimensional force $f(\X_t,t)$. Therefore,
\begin{equation}\label{dSt}
\begin{aligned}
    \frac{dS(t)}{dt} &= \int \sum\limits_{i = 1}^d \left[-f_iJ_i + \frac{J_i^2}{p(\X_t,t)}\right] d\X_t,\\
    &=\pi-\phi,
    \end{aligned}
\end{equation}
where $\pi\equiv\sum_i\int \frac{J_i^2}{p(\X_t,t)}d\X_t$ is non-negative, and is actually the entropy production rate, the rate at which the total entropy of the system and environment changes;
and $\phi\equiv\sum_i\int f_i(\X_t,t)J_i d\X_t$ denotes the entropy flux into or out of the system (from or to) the environment. The entropy flux can be positive, suggesting reduction of the system entropy (a characteristic of emergence of order). Provided that the dynamics reaches equilibrium, both $\pi$ and $\phi$ will vanish, but even if $S(t)$ is stationary, $\pi=\phi\neq0$, which is a key characteristic of nonequilibrium steady states. The above derivations are consistent with those derived in Refs.~\cite{Seifert-2005,Tome-2006}, and are applied to GDMs we study in this paper.

\section{Nonequilibrium physics of reverse generative dynamics}
\subsection{Backward generative SDE}
In this section, we first derive the reverse generative dynamics equation based on the forward diffusion process [Eq.~\eqref{sde0}]. Then we would 
give a thorough physics analysis of this backward generative SDE.

We first define the following backward conditional distribution:
\begin{equation}
P(\X_t,t\vert\X_{t+dt},t+dt)=\frac{P(\X_{t+dt},t+dt\vert\X_t,t)}{p(\X_{t+dt},t+dt)}p(\X_t,t),
\end{equation}
where the Bayes' rule is used. According to Eq.~\eqref{sde0}, we have $\X_{t+dt}-\X_t=f(\X_t,t)dt+\sqrt{2dt}\boldsymbol\eta_t$, where $\boldsymbol\eta_t$ is an i.i.d. Gaussian
random variable of zero mean and unit variance. It is easy to derive that 
\begin{equation}
P(\X_{t+dt},t+dt\vert\X_t,t)\propto\exp\left[-\frac{\left(\X_{t+dt}-\X_t-f(\X_t,t)dt\right)^2}{4dt}\right].
\end{equation}
In addition, the probability ratio
\begin{equation}
\begin{aligned}
\frac{p(\X_t,t)}{p(\X_{t+dt},t+dt)}&=\exp\left[-\left(\ln p(\X_{t+dt},t+dt)-\ln p(\X_t,t)\right)\right]\\
&=\exp\left[-d\X\cdot\nabla\ln p(\X_t,t)-dt\frac{\partial\ln p(\X_t,t)}{\partial t}\right],
\end{aligned}
\end{equation}
where we have done the Taylor expansion of $\ln p(\X+d\X,t+dt)$ in both spatial and temporal dimensions.

Finally, we collect all intermediate results into the Bayes' formula and get
\begin{equation}\label{bayrev}
\begin{aligned}
P(\X_t,t\vert\X_{t+dt},t+dt)&\propto\exp\left[-\frac{\left(\X_{t+dt}-\X_t-f(\X_t,t)dt\right)^2}{4dt}-d\X\cdot\nabla\ln p(\X_t,t)-dt\frac{\partial\ln p(\X_t,t)}{\partial t}\right],\\
&\propto\exp\left[-\frac{\Vert\X_t-\X_{t+dt}+\left(f(\X_t,t)-2\nabla\ln p(\X_t,t)\right)dt\Vert^2}{4dt}\right],
\end{aligned}
\end{equation}
where the order of $\mathcal{O}(dt)$ is neglected. Equation~\eqref{bayrev} suggests that the following SDE for the reverse dynamics reads
\begin{equation}\label{revsde}
\dot\X=f(\X_t,t)-2\nabla\ln p(\X_t,t)+\sqrt{2}\bxi.
\end{equation}
This reverse diffusion equation was first proposed in Ref.~\cite{Anderson-1982}, and can be thought as a non-linear Langevin dynamics, a central subject we shall study in the following sections.

\subsection{Learning the score function}
The gradient of log-likelihood is called the score function in machine learning. It is usually hard to estimate in real data learning, but can be
approximated by a neural network whose parameters are trained to minimize the following mean-squared cost function:
\begin{equation}
\mathcal{L}(\theta)=\mathbb{E}_{\X_t\sim p(\X_t,t)}\left[\Vert s_\theta(\X_t,t)-\nabla\ln p(\X_t,t)\Vert^2\right],
\end{equation}
where $s_\theta$ represents a function implemented by a neural network parameterized by $\theta$. It is proved in a previous work~\cite{Vicent-2011} that the score function can be replaced by $\nabla\ln P(\X_t\vert\X_0)$ (in the sense that the expectation over $p(\X_t,t)$ is considered), which is more convenient to compute as
\begin{equation}\label{smat}
\begin{aligned}
\nabla\ln P(\X_t\vert\X_0)&=-\Sigma^{-1}(\X_t-\X_0e^{-t})\\
&=-\frac{1}{1-e^{-2t}}\I(\X_t-\X_0e^{-t})\\
&=-\frac{1}{\sqrt{1-e^{-2t}}}\mathbf{Z}_t,\\
\end{aligned}
\end{equation}
where $\Sigma=\Sigma_t\I$, and $\mathbf{Z}_t$ is a $d$-dimensional standard Gaussian random variable. Equation~\eqref{smat} can be used to train a neural network in practice. After the score function is learned, the reverse SDE can be used to generate data samples starting from a Gaussian white noise, in a very similar spirit to variational auto-encoder and generative adversarial networks~\cite{DL-2016}.

In our current model, the score function can be computed in an analytic form, which proceeds as follows.
We already know that
\begin{equation}
  p(\X_t,t) = \frac{1}{2}\N\left( \bmu_t,\I \right) + \frac{1}{2}\N\left( -\bmu_t,\I \right),
\end{equation}
where $\bmu_t = \bmu e^{-t}$. We get then the score function~\cite{GMDM-2023}:
\begin{equation}
\begin{aligned}
\nabla \ln p({\X_t},t) & =w_{+, t}(\X_t) \bmu_{+, t}+w_{-, t}(\X_t) \bmu_{-, t}-\X_t \\
&  = w_{+,t}(\X_t)\bmu_t -(1 - w_{+,t}(\X_t))\bmu_t - \X_t \\
&  = (2w_{+,t}(\X_t) - 1)\bmu_t - X_t\\&=\tanh \left( \bmu_t^{ \top }\X_t \right)\bmu_t - \X_t,
\end{aligned}
\end{equation}
where the weight for the positive mean $\bmu_{+,t}$ is given by $w_{+,t}(\X_t) = \frac{1}{{1 + \exp \left( \frac{{\|\X_t - \bmu_t\|^2}}{2} - \frac{{\|\X_t + \bmu_t\|^2}}{2} \right)}} = \frac{1}{{1 + \exp \left( -2\bmu_t^{ \top }\X_t \right)}}$. Note that the score function becomes more complicated but still analytic in the case of non-unit variance of the ground truth Gaussian mixture distribution. We show this complicated expression in Appendix~\ref{app-d}.
Driven by the score function, typical trajectories starting from a standard normal random vectors evolve to target samples of Gaussian mixture data distribution, which is shown as an example in Fig.~\ref{fig1}. What remains mysterious is the nature of the reverse generative process. In the following, we will address this question in our current Gaussian mixture model by using physics concepts.

\subsection{Potential and free energy}\label{pofree}
Next, we follow recent works~\cite{SSB-DM-2023,Free-2023} to introduce the potential and free energy of the reverse SDEs. Note that we formulate these concepts in our current model in the general form and prove their equivalence through a statistical inference mapping of the denoising process, and finally demonstrate how to identify phase transitions during the reverse Langevin dynamics based on our derivation.

We first derive a potential function for the reverse dynamics [Eq.~\eqref{revsde}]. Let $t=T-s$, then decreasing $t$ is equivalent to increasing $s$. The reverse SDE can thus be written as
\begin{equation}\label{revSDE}
\begin{aligned}
d{{\tilde \X}_s} &= \left[ 2\nabla \ln p({{\tilde \X}_s},T - s) - f({\tilde \X}_s,T - s) \right]ds + \sqrt{2}d\mathbf{W}\\
&=  - \nabla U({{\tilde \X}_s},T - s) + \sqrt{2}d\mathbf{W},
\end{aligned}
\end{equation}
where $U({{\tilde \X}_s},T - s)$ is defined as the potential of the reverse dynamics. We have thus the following equality:
\begin{equation}\label{equ}
 - \nabla U({{\tilde \X}_s},T - s) =   2\nabla \ln p({{\tilde \X}_s},T - s) - f({{\tilde \X}_s},T - s).
\end{equation}
Carrying out an integral of both sides in Eq.~\eqref{equ}, we get the expression of the potential as
\begin{equation}
U({{\tilde \X}_s},T - s) =  - 2\ln p({{\tilde \X}_s},T - s) + \int\limits_0^{{{\tilde \X}_s}} {f(Z,T - s)} dZ,
\end{equation}
where we have neglected all irrelevant constants contributed by an arbitrary lower limit (but not ${\tilde X}_s$)  of the integral. In our Gaussian mixture data case, the score function can be exactly computed. Therefore, the potential can be estimated with the following analytic form:
\begin{equation}
U = \frac{1}{2}{{\tilde \X}_s}^2 -2\ln \left[\cosh ({{\tilde \X}_s}\cdot\bmu{e^{ - (T - s)}})\right].
\end{equation}

\begin{figure}
\centering
\includegraphics[width=0.95\textwidth]{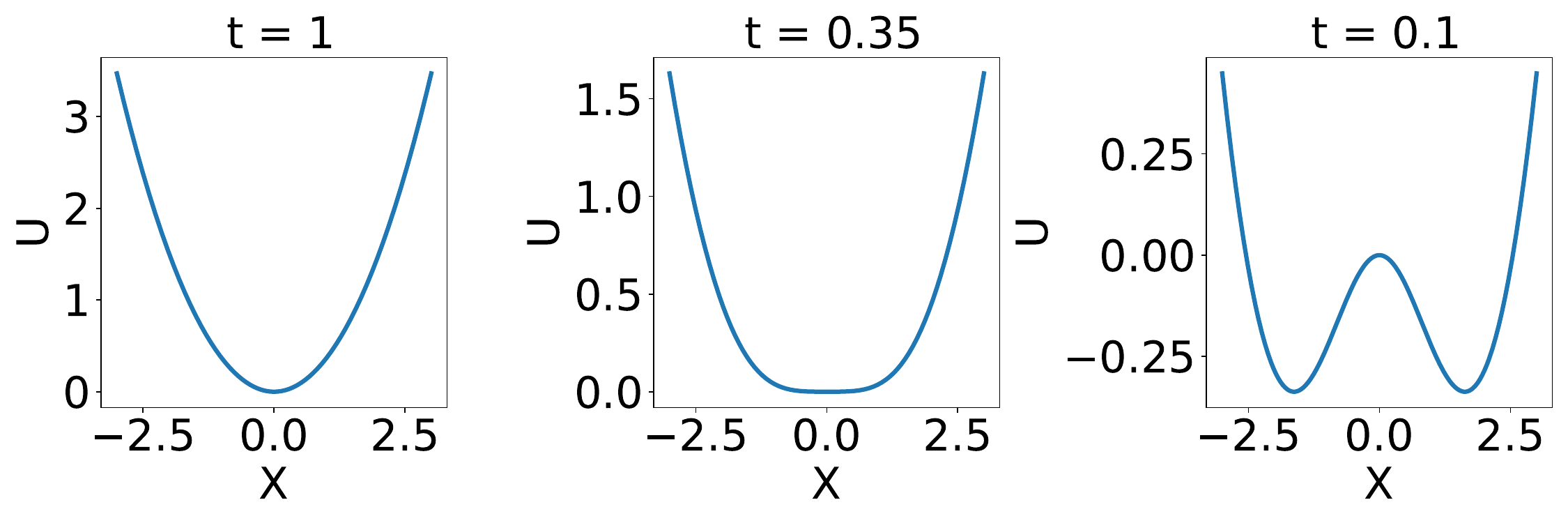}
\caption{Evolution of potential energy  ($\mu = 1$) in one dimension. The speciation transition time is given by $t_S \approx 0.35$.}
\label{potent}
\end{figure}

We reshape the time as $T-s=t$, and then $\tilde{\X}_s=\X_t$, leading to the following form in decreasing $t$:
\begin{equation}\label{potEq}
    U = \frac{1}{2}{\X}^2_t - 2\ln \left[\cosh ({\X_t}\cdot\bmu{e^{ - t}})\right].
\end{equation}
This analytic expression of the potential can be plotted as a function of time and $\X_t$. For an example of one-dimensional dynamics, figure~\ref{potent} shows that at some moment $t=t_S$, the symmetry of the potential is broken, producing two local minima, which bears similarity with the spontaneous symmetry breaking in ferromagnetic Ising model when the temperature is lowered down. This spontaneous symmetry breaking is common in other types of unsupervised learning~\cite{Huang-2019,Huang-2020}. The critical time $t_S$ is thus determined by the condition $\frac{\partial^2U}{\partial X^2}\vert_{X=0}(t_S)=0$, a qualitative change of convexity at $X=0$. This condition leads to the transition condition $\mu^2e^{-2t_S}=\frac{1}{2}$ for our current model setting. The specific moment $t=t_S$ in the reverse dynamics trajectory corresponds to the speciation transition recently identified in Ref.~\cite{Mezard-2024} where an overlap dynamics was analyzed.

Next, we turn to the free energy concept. We assume $p(\X_0)$ as the data distribution, and then according to the forward process, the distribution
of $\X_t$ is expressed as a probability convolution:
\begin{equation}
 p(\X_t,t) = \int \mathcal{N}(\X_t;\X_0e^{-t},\Sigma_t\I)p (\X_0)d\X_0. 
 \end{equation}
The score function can then be expressed as 
\begin{equation}
\begin{aligned}
\nabla \ln p(\X_t,t)&=\Sigma_t^{-1}\left[\frac{\int d\X_0p(\X_0)\X_0e^{-t}\mathcal{N}(\X_t;\X_0e^{-t},\Sigma_t\I)}{\int d\X_0p(\X_0)\mathcal{N}(\X_t;\X_0e^{-t},\Sigma_t\I)}-\X_t\right]\\
& =\Sigma_t^{-1} \left[\lrbka{\X_0e^{-t}}_{\X_t} - \X_t\right].
\end{aligned}
\end{equation}
This form of score function suggests that a statistical inference can be defined below~\cite{Lenka-2023}:
\begin{equation}\label{smeq}
\begin{aligned}
    P(\X_0\vert \X_t,t)& = \frac{P(\X_t,t|\X_0)p(\X_0)}{p(\X_t,t)}\\
    &=\frac{\mathcal{N}(\X_t;\X_0e^{-t},\Sigma_t\I)p(\X_0)}{p(\X_t,t)}\\
       &  \propto \exp \left[ -\frac{1}{2} (\X_t-\X_0e^{-t})^\top\Sigma^{-1} (\X_t-\X_0e^{-t}) + \ln p(\X_0)\right],
\end{aligned}
\end{equation}
where $\Sigma=\Sigma_t\I$.
We can thus write down an equivalent Hamiltonian $\mathcal{H}(\X_0\vert \X_t,t)=\frac{1}{2}(\X_t-\X_0e^{-t})^\top\Sigma^{-1} (\X_t-\X_0e^{-t}) - \ln p(\X_0)$, and 
the partition function reads
\begin{equation}
    Z(\X_t,t) = \int e^{ - \mathcal{H}(\X_0\vert \X_t,t)} d\X_0.
\end{equation}
Note that this is a generalized form of that introduced in Ref.~\cite{Free-2023}. The dynamic state $\X_t$ during the reverse dynamics plays the role of quenched disorder in
traditional spin glass theory~\cite{Huang-2022}.

\begin{figure}
\centering
\includegraphics[width=0.8\textwidth]{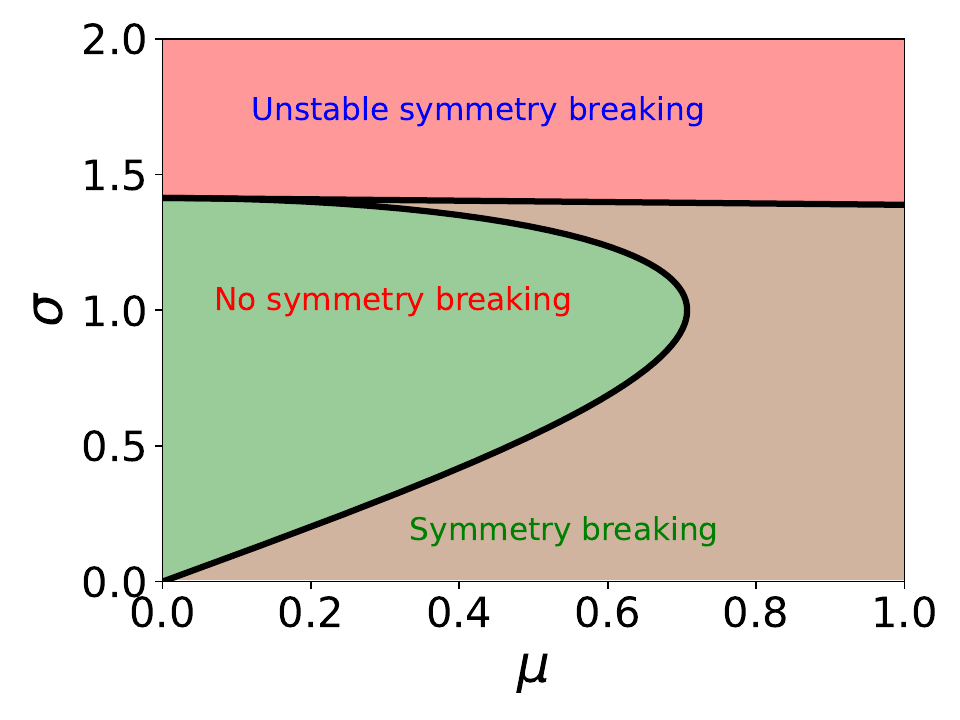}
\caption{Phase diagram of symmetry breaking in one-dimensional example of arbitrary data mean $\mu$ and variance $\sigma^2$. The concrete examples of three phases are shown in Fig.~\ref{Typ}.}
\label{phase}
\end{figure}

\begin{figure}
\centering
\includegraphics[width=0.8\textwidth]{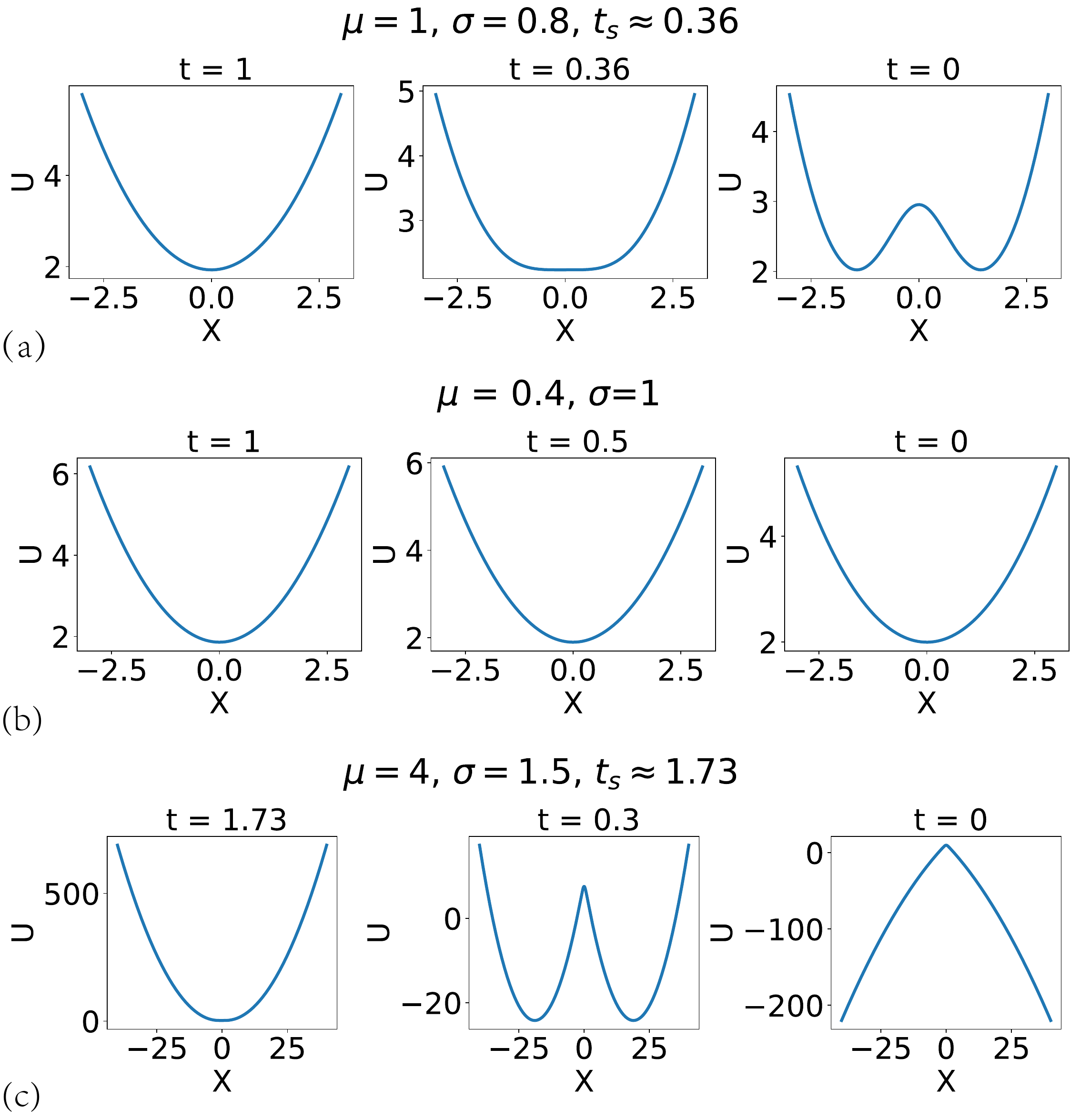}
\caption{Typical profiles of potential in the reverse generative process corresponding to different phases in Fig.~\ref{phase}. (a) Stable symmetry breaking.
(b) Symmetry un-breaking. (c) Unstable symmetry breaking. }
\label{Typ}
\end{figure}

The free energy can then be written as $F(\X_t,t)=-\frac{1}{\beta}\ln Z(\X_t,t)$, where an equivalent inverse temperature $\beta=\Sigma_t^{-1}$ (different from that introduced 
in Ref.~\cite{Free-2023}). It is straightforward to compute
the following free-energy gradient:
\begin{equation}
\begin{aligned}
    \nabla F(\X_t,t) &= \nabla \left( - \frac{1}{\beta }\ln Z(\X_t,t)\right)\\
    & =  - \frac{1}{\beta }\frac{{\nabla Z(\X_t,t)}}{{Z(\X_t,t)}}\\
    & =  - \lrbka{\X_0e^{-t}}_{\X_t}+\X_t,
\end{aligned}
\end{equation}
which is exactly the gradient of log-state-likelihood (up to a pre-factor), i.e., $\nabla F=-\frac{1}{\beta}\nabla\ln p$, from which we can finally derive the explicit
form of the free energy:
\begin{equation}
F=-\frac{1}{\beta}\ln\cosh(\X_t\cdot\bmu_t)+\frac{1}{2\beta}\X_t^2.
\end{equation}
It is then convenient to write the driving force in the reverse dynamics [see Eq.~\eqref{revSDE}], i.e., $2\nabla\ln p-f=-\nabla\tilde{F}$, and then it is straightforward to derive the following expression for the generalized free energy $\tilde F$:
\begin{equation}
\tilde{F}=2\beta F+\int fd\tilde{\X}_s,
\end{equation}
which is exactly the potential energy derived before [Eq.~\eqref{potEq}]. This suggests that, the reverse generative dynamics can be considered as minimizing the generalized free energy in the dynamic-state space, and the generated samples follows the minimum free energy principle. We thus conclude that the speciation transition corresponds to a change of curvature of the free energy landscape. Although our results are derived in the high-dimensional Gaussian mixture context and formulated as a statistical inference, the conclusion is consistent with
that derived in a simpler setting through a regularized free energy~\cite{Free-2023}.

\blue{We finally discuss the symmetry breaking transition in arbitrary values of data mean $\mu$ and variance $\sigma^2$. For simplicity, we focus on one-dimensional example. According to the theoretical analysis in this section, the speciation time $t_S$ is determined by $t_S=\frac{1}{2}\ln\left(\mu^2+\sqrt{\sigma^4-2\sigma^2+\mu^4+1}\right)$ where the potential starts the spontaneous symmetry breaking, which is obtained by searching for the time when the second derivative of the potential or free energy vanishes (see detailed calculations in Appendix~\ref{app-bb}). The theoretical formula based on a perturbative analysis in Ref.~\cite{Mezard-2024} does not apply to our current case where $d=1$ and speciation may take place at a small time. The following finding is also not revealed in a recent work~\cite{SSB-DM-2023} where the arbitrary variance was not considered. Hence, in our current setting, we reveal a new type of phase diagram for this simple setting (Fig.~\ref{phase}). More precisely, $\mu^2+\sqrt{\sigma^4-2\sigma^2+\mu^4+1}=1$ determining a phase boundary between symmetry breaking and symmetry un-breaking. In the symmetry un-breaking phase, we do not have a positive value of $t_S$. Moreover, it is surprising that when the variance is higher than a critical value, we have another phase transition to an unstable symmetry breaking, whose threshold is determined by $t_{USB}=\frac{1}{2}\ln(\sigma^2-1)$ (see details in Appendix~\ref{app-bb}). If the unstable symmetry breaking is triggered, i.e., $t_{USB}\geq0$, the symmetry breaking profile of double-minima potential is destabilized to an inverted U shape (Fig.~\ref{Typ}). The slope of the potential at $X_t=\infty$ changes discontinuously to a positive value, since at the threshold, the slope is still negative. An immediate result is that the generative sampling fails for the reverse diffusion process. Although we clarify this phase diagram in the simple one-dimensional diffusion process, it is still very interesting to see whether the picture qualitatively holds in high dimensional generative process, which is more relevant in practical applications. We leave this idea in future works.}

\subsection{Fluctuation theorem in the reverse dynamics}
We have derived the following reverse SDE [see Eq.~\eqref{revsde}]:
\begin{equation}
{\rm{d}}{\X_t} =  - 2\tanh \left( {\bmu _t^{\T}{\X_t}} \right)\bmu _t{\rm{d}}t + {\X_t}{\rm{d}}t + \sqrt 2 {\rm{d}}\mathbf{W}.
\end{equation}
With increasing time $s$, we make a definition $\tilde{\X}(s)=\X(t)$ as before, where $s=T-t$, and we have the following 
equivalent SDE:
\begin{equation}\label{back}
d{{\tilde \X}_s} = \left[ {2\tanh \left( {\bmu _{T - s}^{\T}{{\tilde \X}_s}} \right)\bmu _{T - s} - {{\tilde \X}_s}} \right]ds + \sqrt 2 d\mathbf{W}.
\end{equation}
We then use the $\lambda$-convention for the following discretization:
\begin{equation}
\frac{{{{\tilde \X}_{s + ds}} - {{\tilde \X}_s}}}{{ds}} = \left[ {2\tanh \left( \bmu _{T - s - \lambda ds}^{\T}\{\lambda{{\tilde \X}_{s + ds}} +(1-\lambda) {{\tilde \X}_s}\} \right)\bmu _{T - s - \lambda ds} - (\lambda{{\tilde \X}_{s + ds}} +(1-\lambda) {{\tilde \X}_s})} \right] + \sqrt 2 \boldsymbol\eta_s, 
\end{equation}
where $\boldsymbol\eta_s \sim \N(0,\frac{1}{{ds}}\I)$. The conditional probability of $\tilde{\X}_{s+ds}$ can be expressed using the noise distribution via the following probability density transform:
\begin{equation}
\begin{aligned}
  &P\left( {{{\tilde \X}_{s + ds}}|{{\tilde \X}_s}} \right) = P\left( {\sqrt 2 \boldsymbol\eta } \right)\left| {\frac{{\partial \sqrt 2\boldsymbol \eta }}{{\partial {{\tilde \X}_{s + ds}}}}} \right|\\
  &   \propto \exp \left[ - \frac{{{\left({{\dot {\tilde \X}}_s} - 2\tanh (\bmu _{T - s}^{\T}{{\tilde \X}_s})\bmu _{T - s} + {{\tilde \X}_s}\right)^2}ds}}{4}\right.\\
  &\left. - \lambda \nabla  \cdot \left(2\tanh (\bmu _{T - s}^{\T}{{\tilde \X}_s})\bmu _{T - s} - {{\tilde \X}_s}\right)\right],
\end{aligned}
\end{equation}
where $\left| {\frac{{\partial \sqrt 2 \boldsymbol\eta }}{{\partial {{\tilde \X}_{s+ds }}}}} \right|$ represents the Jacobian.
    Finally, we use the Markovian chain property and get
    \begin{equation}
\begin{aligned}
   &P\left( {\tilde X([T] )|{{\tilde \X}_0}} \right) = \prod\limits_s {P\left( {{{\tilde \X}_{s + ds}}|{{\tilde \X}_s}} \right)} \\&    \propto {e^{ - \int_0^T d s\mathop  \odot \limits^\lambda  \left[ {\frac{{{\left(\dot {\tilde \X} - 2\tanh (\bmu _{T - s}^{\T}\tilde \X)\bmu _{T - s} + \tilde \X\right)^2}}}{4} + \lambda \nabla  \cdot [2\tanh (\bmu _{T - s}^{\T}\tilde \X)\bmu _{T - s} - \tilde \X]} \right]}}.\label{mavb}
   \end{aligned}
   \end{equation}

Given an initial condition, the path probability of a trajectory $\tilde X([T])$ with time from $0$ to $T$ can be represented as
\begin{equation}
    P(\tilde X([T] )|\tilde \X_0) = \N\exp [ - \A(\tilde X([T] ))],
    \end{equation}
    where $\N$ denotes a normalization, and
    \begin{equation}
    \A(\tilde X([T] )) = \int_0^T d s\mathop  \odot \limits^\lambda  \left[\frac{{{\left(\dot {\tilde \X} - 2\tanh (\bmu _{T - s}^{\T}\tilde \X)\bmu _{T - s} + \tilde \X\right)^2}}}{4}
  + \lambda \nabla  \cdot [2\tanh \left( {\bmu _{T - s}^{\T}\tilde \X} \right)\bmu _{T - s} - \tilde \X]\right].
\end{equation}
Taking a backward process on the same forward trajectory, we have $\X(t)=\X(T-s)=\tilde{\X}(s)$, and thus we have
$\dot{\X}(t)=\dot{\X}(T-s)=-\dot{\tilde{\X}}(s)$. The path probability for this reverse version of the forward trajectory can be derived 
similarly based on the new definition of the state variable. The resulting formula is given below.
\begin{equation}
    P(X([T] )|\X_0) = \N\exp [ - \A(X([T]))],
    \end{equation}
    where
    \begin{equation}
     \A(X([T] )) = \int_0^T d s\mathop  \odot \limits^{1 - \lambda }  \left[\frac{{{\left(-\dot{\tilde{\X}} - 2\tanh (\bmu _{T - s}^{\T}{{\tilde \X}})\bmu _{T - s} + {{\tilde \X}}\right)^2}}}{4}
    + \lambda \nabla  \cdot [2\tanh \left( {\bmu _{T - s}^{\T}\tilde \X} \right)\bmu _{T - s}- \tilde \X]\right].
\end{equation}

\begin{figure}
\centering
\includegraphics[width=0.7\textwidth]{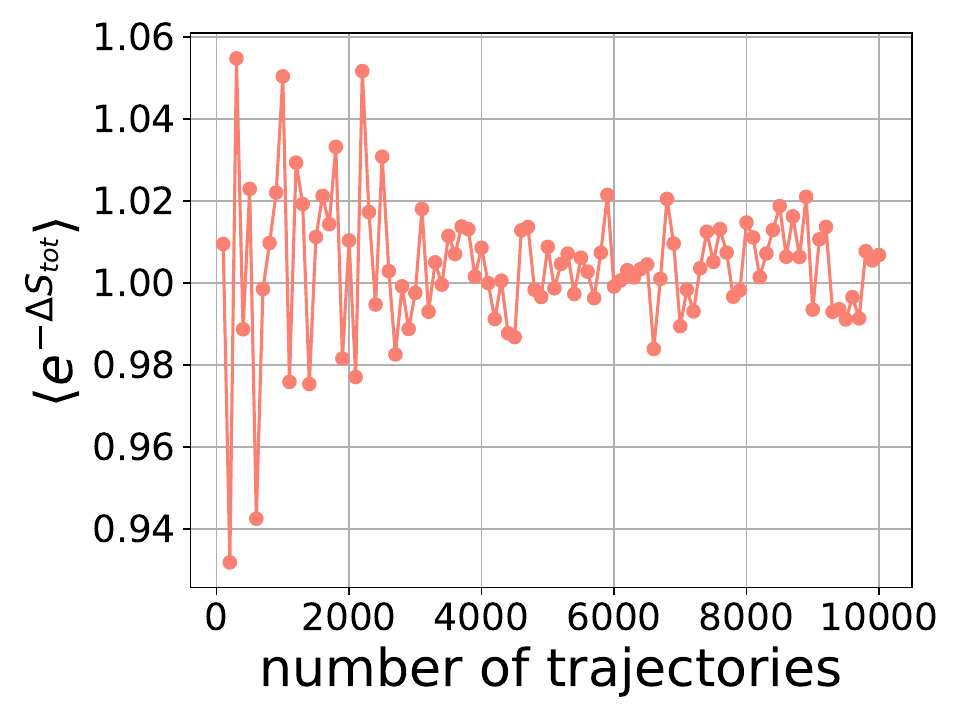}
\caption{Ensemble average of $e^{ - \Delta {S_{tot}}}$ as a function of the number of trajectories used to estimate the average, simulated by solving the backward SDE. $\mu = 1$, and other parameters are detailed in Appendix~\ref{app-a}.}
\label{backwardS}
\end{figure}

\begin{figure}
\centering
\includegraphics[width=0.85\textwidth]{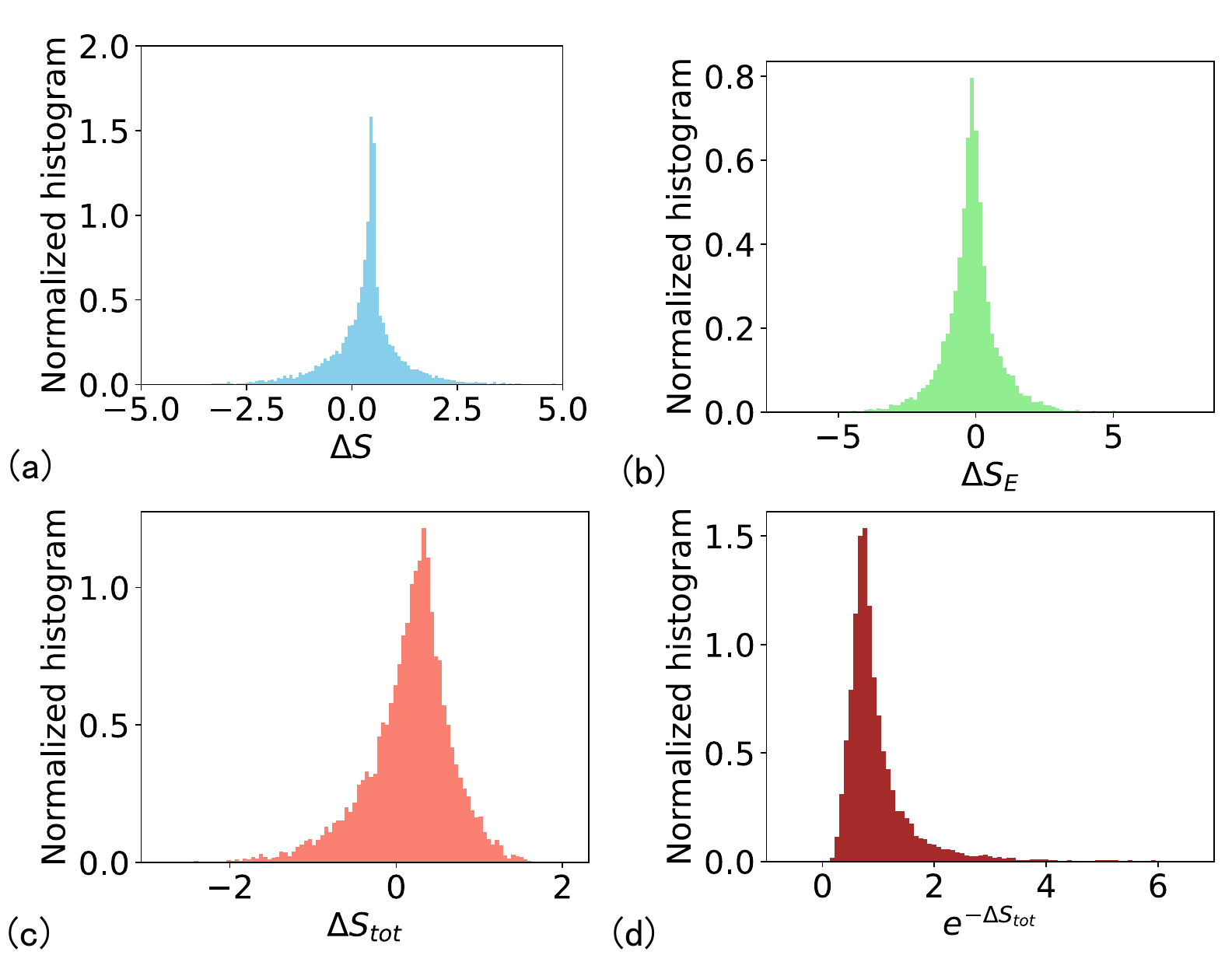}
\caption{ Normalized histogram of entropy changes estimated from $10\,000$ Trajectories for the reverse generative process. $\mu = 1$, and other parameters are detailed in Appendix~\ref{app-a}. (a) Statistics of system entropy change. (b) Statistics of environment entropy change. (c) Statistics of total entropy change. (d) Statistics of $e^{-\Delta S_{\text {tot }}}$.}
\label{backward}
\end{figure}

The entropy change in the environment is thus calculated as follows,
\begin{equation}
\begin{aligned}
    \Delta {S_E} &= \ln \left[\frac{{P(\tilde X([T] )|\tilde \X_0)}}{{P(X([T] )|\X_0)}}\right]\\
    & = \int_0^T ds\mathop  \odot \limits^{\frac{1}{2}}  \left[ \dot{\tilde{\X}} \cdot(2\tanh ({\bmu _{T - s}^{\T}} \tilde{\X}) \bmu_{T - s} - \tilde{\X}) \right]\label{smb}.
\end{aligned}
\end{equation}
The change of the system entropy is calculated as the ratio between the probabilities of the initial and final states. More precisely,
\begin{equation}
\begin{aligned}
  \Delta S & = \ln \left[\frac{p( {\tilde \X_0})}{p(\tilde \X_T)}\right]\\&   = \ln \left[ {\frac{{\exp ( - \frac{{{{(\tilde \X_0 - \bmu {e^{ - T}})}^2}}}{2}) + \exp ( - \frac{{{{(\tilde \X_0 + \bmu{e^{ - T}})}^2}}}{2})}}{{\exp ( - \frac{{{{(\tilde \X_T - \bmu)}^2}}}{2}) + \exp ( - \frac{{{{(\tilde \X_T + \bmu)}^2}}}{2})}}} \right].\label{sb}
\end{aligned}
\end{equation}
Combining the above two entropy contributions, we get the total entropy change:
\begin{equation}
    \Delta {S_{tot}} = \Delta {S_E} + \Delta S = \ln \left[\frac{{P\left[ {\tilde X([T])} \right]}}{{P\left[ {X([T] )} \right]}}\right],
\end{equation}
where $\tilde X([T])$ and $X([T])$ have included the starting point.
Therefore, the reverse dynamics also obeys the following integral fluctuation theorem:
\begin{equation}
\begin{aligned}
    \left\langle {{e^{ - \Delta {S_{tot}}}}} \right\rangle  & = \int {P\left[ {\tilde X([T] )} \right]{e^{ - \Delta {S_{tot}}}}d\tilde X([T])}  \label{equ}\\
    &  = \int {P\left[ {X([T] )} \right]dX([T] )}  \\
    & = 1,
\end{aligned}
\end{equation}
which suggests that the second law of thermodynamics $ \lrbka{\Delta S_{tot}}\ge 0$. This integral fluctuation theorem is verified by our experiments in Fig.~\ref{backwardS} of one-dimensional example. Detailed contribution of entropy is given in Fig.~\ref{backward} as well. Compared to the forward process, the system and total entropy changes are 
biased toward positive values. Experimental details are given in Appendix~\ref{app-a}.

\subsection{Entropy production rate}
In this section, we analyze the time-dependent entropy change in both forward and reverse dynamics, focusing on
the reverse generative dynamics.

For the ensemble entropy production rate, we consider the following initial distribution:
\begin{equation}
 p({\X_0}) = \frac{1}{2}\N\left( {{\bmu },{\sigma ^2}\I} \right) + \frac{1}{2}\N\left( { - {\bmu },{\sigma ^2}\I} \right),
 \end{equation}
 where all components of $\bmu$ are equal to a positive value $\mu$.
This implies that the probability of $\X_t$ at time $t$ is given by a convolution of the initial distribution and the Gaussian transition
kernel:
\begin{equation}\label{pxt}
p({\X_t},t)  = \frac{1}{2}\N\left( {\bmu _t,(\I - {e^{ - 2t}}\I) + {\sigma ^2\I}{e^{ - 2t}}} \right)
    + \frac{1}{2}\N\left( { - \bmu _t \I,(\I - {e^{ - 2t}}\I) + {\sigma ^2\I}{e^{ - 2t}}} \right),
\end{equation}
where $\bmu_t=\bmu e^{-t}$. 

Next, we write down the expression of the probability current for both forward and reverse dynamics:
\begin{equation}
\J=f(\X_t,t)p(\X_t,t)-\nabla p(\X_t,t),
\end{equation}
where $f(\X_t,t)=-\X_t$ for the forward dynamics, and $2\tanh(\bmu_{T-s}^\T\tilde{\X}_s)\bmu_{T-s}-\tilde{\X}_s$ for the reverse dynamics ($\sigma^2=1$). Note that the forward and backward SDEs share the same state probability distribution $p(\X_t,t)$ (see Eq.~\eqref{pxt}, and a proof given in Appendix~\ref{app-b}). Note also that the probability current of the forward OU process has the same magnitude but opposite direction with the reverse generative SDE (see a proof in Appendix~\ref{app-c}). Inserting the above definitions, one can estimate the entropy production rate $\pi$ and the entropy flux $\phi$ according to Eq.~\eqref{dSt}.

Although our formula applies to the high-dimensional case (the dimensionality can be set to an arbitrary number), we consider a one-dimensional example to analyze the entropy production rate and entropy flux for simplicity. Due to the anti-symmetry properties of probability currents and the identical state probability, we have the same entropy production rate:
\begin{equation}
\pi  ={\pi ^ {*} }= \int {\frac{{{{\left[{X_t}p({X_t},t) + \frac{{\partial p({X_t},t)}}{{\partial {X_t}}}\right]}^2}}}{{p({X_t},t)}}dX_t} .
\end{equation}
We use the superscript $*$ to indicate the reverse process.
However, the entropy fluxes are different depending on the specific forms of the drift force. According to Sec.~\ref{eep}, we have the following results:
\begin{equation}
\begin{aligned}
 \phi &= \int {{X_t}\left[{X_t}p({X_t},t) + \frac{{\partial p({X_t},t)}}{{\partial {X_t}}}\right]dX_t} ,\\
    \phi^{*}  &  = -\int {f_{rev}(X_t,t)\left({X_t}p({X_t},t) + \frac{{\partial p({X_t},t)}}{{\partial {X_t}}}\right)dX_t}.
   \end{aligned}
\end{equation}
Note that for $\sigma^2=1$, the drift force for the reverse dynamics is given by $f_{rev}(X_t,t)={2\tanh \left( {\mu _t{X_t}} \right)\mu _t - {X_t}}$. The results for the case of $\sigma^2\neq1$ are derived in Appendix~\ref{app-d}.

\begin{figure}
\centering
\includegraphics[width=0.8\textwidth]{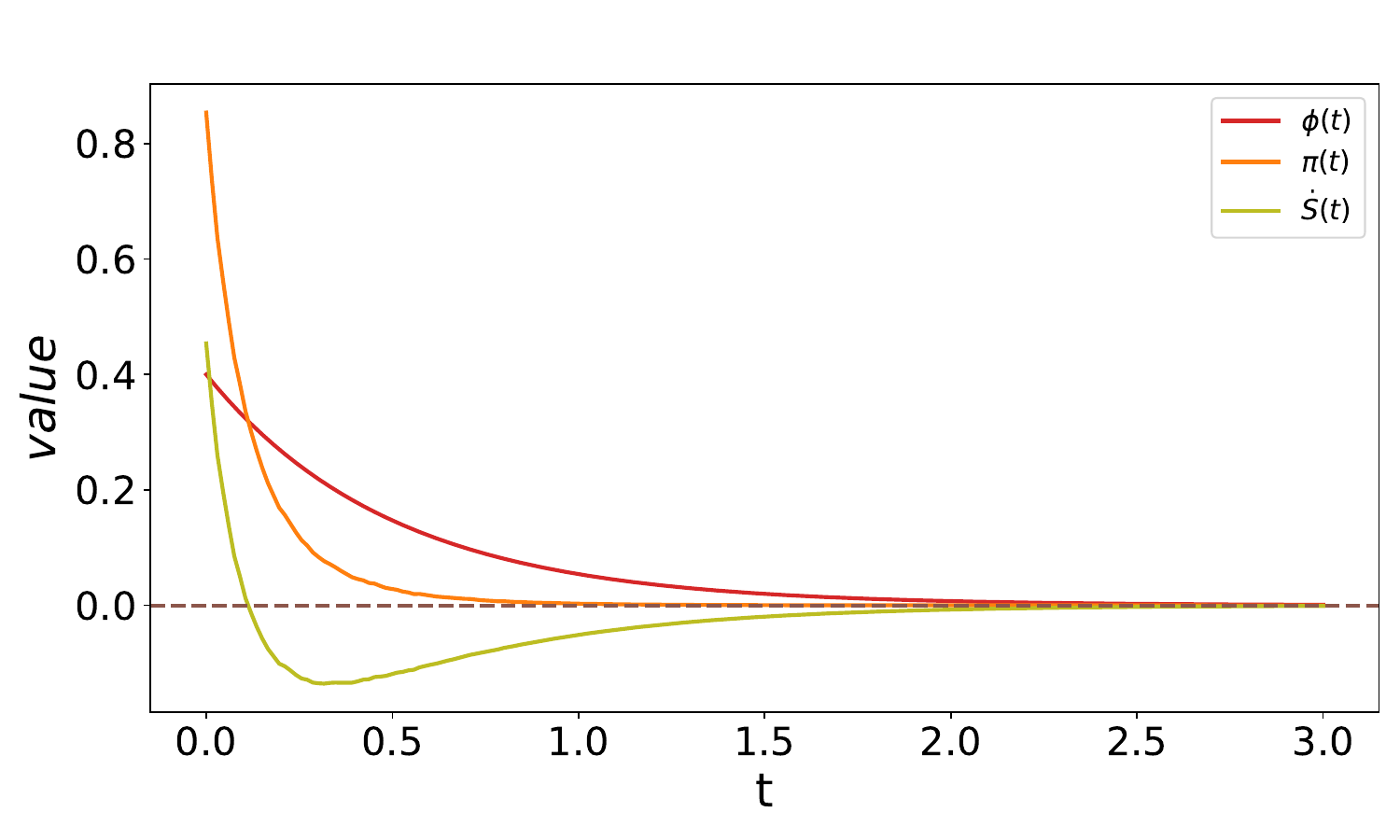}
\caption{Entropy production rate and flux ($\mu =1$, $\sigma^2= 0.4$).  The results are plotted for the forward OU process from time $0$ to $3$.
 }
\label{fig7}
\end{figure}

\begin{figure}
\centering
\includegraphics[width=0.8\textwidth]{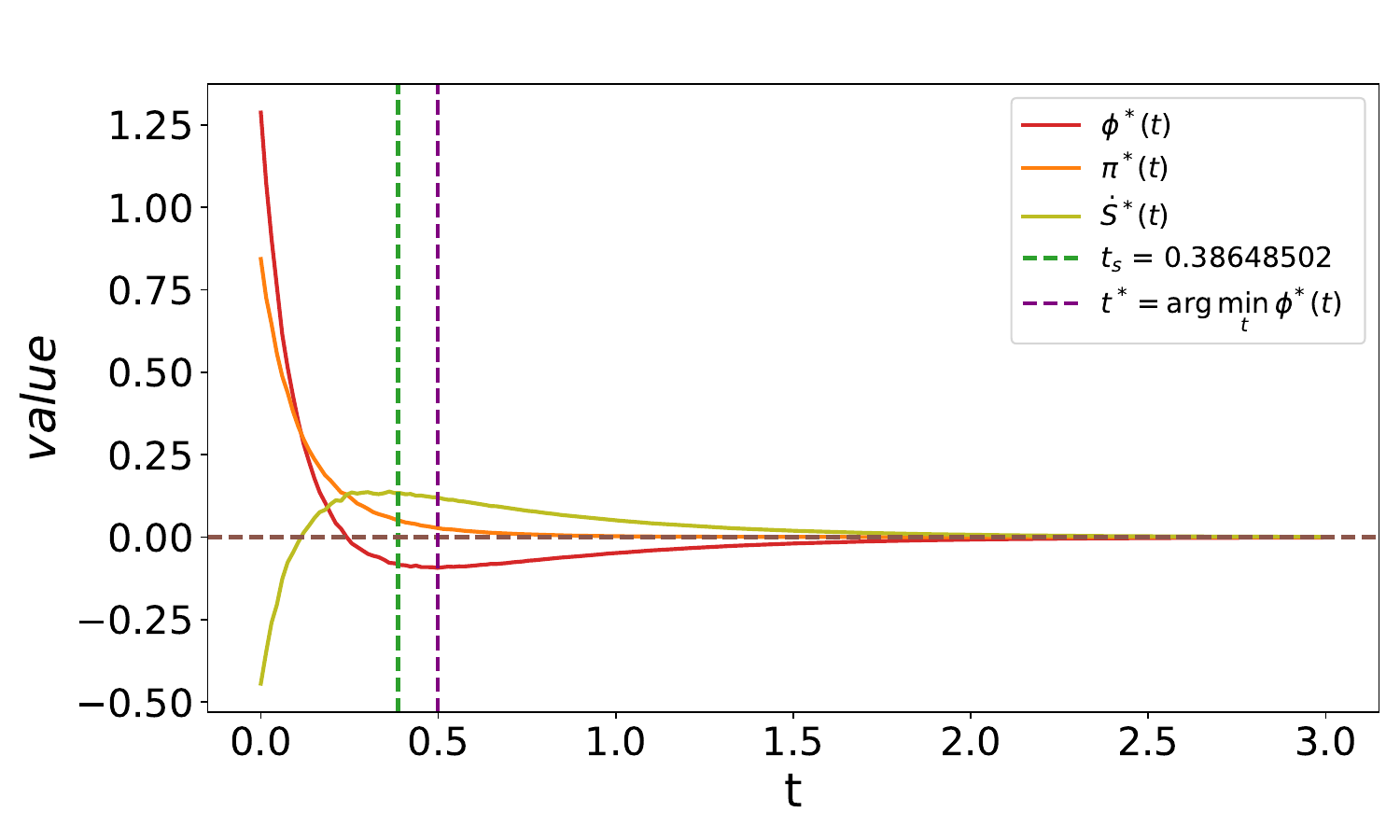}
\caption{Entropy production rate and flux
 ($\mu =1$, $\sigma^2 = 0.4$).  The results are plotted for the reverse generative diffusion process from time $3$ to $0$.
 }
\label{fig8}
\end{figure}

\blue{As derived in Sec.~\ref{pofree}, the speciation time $t_S=\frac{1}{2}\ln\left(\mu^2+\sqrt{\sigma^4-2\sigma^2+\mu^4+1}\right)$ for a one-dimensional generative examples of arbitrary values of mean and variance.} This corresponds exactly to the first appearance of the inflection point on the potential or free energy curve (see Sec.~\ref{pofree}). As shown in Fig.~\ref{fig7} for the forward OU process, the entropy production rate drops monotonically toward zero, while the system entropy rate decreases first and then increases toward zero. With increasing time in a reverse process (from noise sample to data sample), the entropy flux is first negative, and after some time step earlier than $t_S$, the flux becomes positive, indicating that the order is generated (Fig.~\ref{fig8}). The change rate of system entropy is first positive, and particularly before $t_S$, the system entropy rate starts to decrease. In particular, the rate drops sharply as the starting point ($t=0$) is approached, and at $t=0$ one real sample is generated. Meanwhile, the entropy flux increases sharply as well, indicating a generative process in
sampling the target data space.

\subsection{Glass transition}
It was recently argued that at a time less than $t_S$ (in the reverse dynamics), a collapse transition would take place, i.e., the trajectory condenses onto one single sample of data distribution~\cite{Mezard-2024}, which further clarified that the essence is the glass transition. This recent analysis relies on the empirical distribution of time-dependent state (e.g., $\X_t$ here), where $n$ data points are selected as an initial condition for the diffusion dynamics. Here, we re-interpret this picture using the Franz-Parisi potential, a powerful statistical physics tool to characterize the geometric structure of glassy energy landscape~\cite{FP-1998,Huang-2014}, as in our current case, the score can be analytically computed, and as shown before, the dynamical state at some specific moment during the reverse process plays a role of quenched disorder for the statistical inference of the denoising process.

We start from the statistical inference defined by
Eq.~\eqref{smeq}. We select an equilibrium reference configuration, namely $\X'_0$, and consider a restricted Boltzmann measure
\begin{equation}
p(\X_0\vert\X_t,\X'_0)=\frac{1}{Z(\X_t,\X'_0,q)}\exp\left[-\mathcal{H}(\X_0\vert\X_t,t)\right]\delta\left[q-d(\X_0,\X'_0)\right],
\end{equation}
where $d(\X_0,\X'_0)$ denotes the Euclidean distance of two high dimensional vectors. This restricted Boltzmann measure can be transformed to a soft constraint with a coupling filed $\epsilon$. Therefore, we shall focus on the following constrained free energy:
\begin{equation}
F(\epsilon)=-\left\langle\int d\X'_0p(\X'_0\vert\X_t)\ln\int d\X_0e^{-\mathcal{H}(\X_0\vert\X_t,t)+\epsilon d(\X_0,\X'_0)}\right\rangle_{\X_t},
\end{equation}
where $\X_t\sim p(\X_t,t)$ derived before, and acts as quenched disorder as in usual spin glass theory~\cite{Mezard-1987}. We thus define the Franz-Parisi potential $V(q)=\lrbka{-\ln Z(\X_t,\X'_0,q)}$, where $\lrbka{\cdot}$ means the average over $\X'_0$ and $\X_t$. The Franz-Parisi potential $V(q)$ is obtained by a Legendre transform of $F(\epsilon)$  which is given below.
\begin{equation}
V(q)=\min_{\epsilon} F(\epsilon)+\epsilon q,
\end{equation}
where $q$ is determined by $\frac{\partial F}{\partial\epsilon}=-q$.

The Franz-Parisi potential develops the second minimum if a dynamical glass transition occurs, implying that the ergodicity breaks. Furthermore, if the second minimum reaches the same height with the first minimum, a static glass transition occurs with vanishing complexity~\cite{FP-1998}. 

We next show an example of one-dimensional Franz-Parisi potential which can be computed directly. The Hamiltonian reads,
\begin{equation}
    {\cal H}({X_0}|{X_t},t) = \frac{{{{({X_t} - {X_0}{e^{ - t}})}^2}}}{{2(1 - {e^{ - 2t}})}} - \ln \left[\frac{1}{{2\sqrt {2\pi {\sigma ^2}} }}\exp ( - \frac{{{{({X_0} - \mu )}^2}}}{{2{\sigma ^2}}}) + \frac{1}{{2\sqrt {2\pi {\sigma ^2}} }}\exp ( - \frac{{{{({X_0} + \mu )}^2}}}{{2{\sigma ^2}}})\right].
\end{equation}
The constrained partition function can be written as follows,
\begin{equation}
Z({X_t},{X_0}' ,q,t) = \int {\exp ( - {\cal H}({X_0}|{X_t},t))\delta (q - {{({X_0} - {X_0}')}^2})d{X_0}} \label{partition}.
\end{equation}
Therefore, the Franz-Parisi potential reads,
\begin{equation}
    V(q,t) =  - {\left\langle {\int {p\left( {{X_0}' \mid {X_t},t} \right)\ln Z({X_t},{X_0}' ,q,t)} d{X_0}^\prime } \right\rangle _{{X_t}}}.
\end{equation}
To proceed, we have to use the following property of Dirac delta function:
\begin{equation}
    \delta (q - {({X_0} - {X_0}' )^2}) = \frac{1}{{2\sqrt q }}\left[\delta ({X_0} - {X_0}'  - \sqrt q ) + \delta ({X_0} - {X_0}'  + \sqrt q )\right].
\end{equation}
Then, we can simplify the constrained partition function as follows.
\begin{equation}
\begin{aligned}
Z({X_t},{X_0}' ,q,t) &= \int {\exp ( - {\cal H}({X_0}|{X_t},t))\delta (q - {{({X_0} - {X_0}' )}^2})d{X_0}} \\& = \frac{1}{{2\sqrt q }}\exp ( - H({X_0}'  + \sqrt q |{X_t},t)) + \frac{1}{{2\sqrt q }}\exp ( - H({X_0}'  - \sqrt q |{X_t},t)).
\end{aligned}
\end{equation}
We already know the joint distribution $p(X_0',X_t,t)$ as follows,
\begin{equation}
\begin{aligned}
    &p\left( {{X_0}' ,{X_t},t} \right)  = p(X_t,t\vert X_0')p(X_0')\\
    &=\frac{{\exp ( - \frac{{{{({X_t} - {X_0}' {e^{ - t}})}^2}}}{{2(1 - {e^{ - 2t}})}})}}{{\sqrt {2\pi (1 - {e^{ - 2t}})} }}\left( {\frac{1}{{2\sqrt {2\pi {\sigma ^2}} }}\exp ( - \frac{{{{({X_0}'  - \mu )}^2}}}{{2{\sigma ^2}}}) + \frac{1}{{2\sqrt {2\pi {\sigma ^2}} }}\exp ( - \frac{{{{({X_0}^\prime  + \mu )}^2}}}{{2{\sigma ^2}}})} \right)\\
    &=\frac{1}{{4\pi \sqrt {{\sigma ^2}(1 - {e^{ - 2t}})} }} {\exp ( - \frac{{{{({X_t} - {X_0}' {e^{ - t}})}^2}}}{{2(1 - {e^{ - 2t}})}})\left[\exp ( - \frac{{{{({X_0}'  - \mu )}^2}}}{{2{\sigma ^2}}}) + \exp ( - \frac{{{{({X_0}'  + \mu )}^2}}}{{2{\sigma ^2}}})\right]} .
\end{aligned}
\end{equation}
It is therefore straightforward to use Monte-Carlo method to estimate the Franz-Parisi potential. We first generate $\mathcal{T}$ pairs of $(X_0',X_t)$ according to
the above joint distribution. Then the potential is estimated in a simple way.
\begin{equation}
    V(q,t) =  - \frac{1}{\mathcal{T}}\sum\limits_{\ell= 1}^\mathcal{T} \ln Z(X_{t,\ell},X_{0,\ell}' ,q,t).
\end{equation}

\begin{figure}[ht]
\centering
\includegraphics[width=0.9\textwidth]{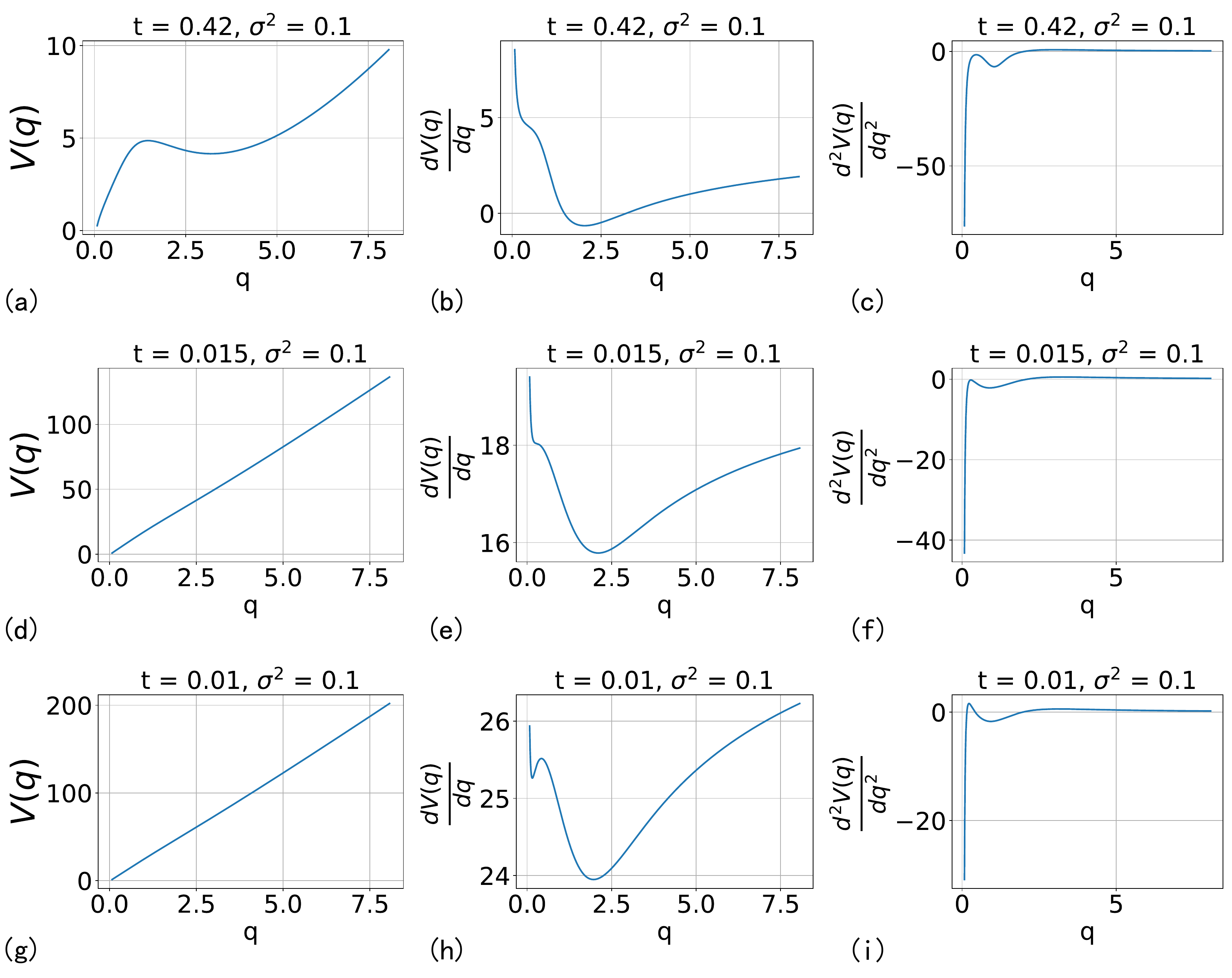}
\caption{The Franz-Parisi potential and its first and second derivatives for the case of $\mu = 1$ and $\sigma^2 = 0.1$. $t_S=0.42622$. Three time steps are considered, i.e., $ t = 0.42, t = 0.015$, and $ t = 0.01$.}
\label{FP-fig}
\end{figure}

The Franz-Parisi potential is plotted as a function of Euclidean distance $q$ for the one-dimensional example in Fig.~\ref{FP-fig}. When the time decreases from the starting point of the reverse diffusion, the potential develops an inflection point where the second derivative vanishes at some moment. As the time approaches the starting time ($t=0$), more inflection points appear, due to fragmented (inferred) data space given the current $X_t$ value. \blue{This demonstrates that the reverse trajectory will collapse onto a single data sample}. The complicated yet tractable computation of the high dimensional case is left for future works, where order parameters characterizing the transition would emerge.

\section{Conclusion}
In this paper, we provide a thorough study of generative diffusion models widely used in unsupervised machine learning, especially in Sora~\cite{Sora-2024}. We use nonequlibrium physics concepts to dissect the mechanisms of the diffusion model, and derive the entropy production, the second law of stochastic thermodynamics and path probability, and we also treat the reverse generative process as a statistical inference problem where the state variable at the reverse time step serves as quenched disorder like in a standard spin glass problem~\cite{Huang-2022}, and apply the concept of equilibrium physics such as potential energy, free energy, and Franz-Parisi potential to study different kinds of phase transitions in the reverse process. Although some results are revealed by recent works from different angles~\cite{SSB-DM-2023,Free-2023,Mezard-2024,Path-2024}, our results provide a complete physics picture from nonequilibrium (such as fluctuation theorem, entropy production) to equilibrium (especially spontaneous symmetry breaking and geometric method to study glass transition), making generative diffusion models more transparent. Our main contributions have been elaborated in the introduction. We hope these methods or concepts elaborated in our paper will inspire statistical physicists to study the currently active research frontier of diffusion models, for which a pure understanding may inspire new inductive biases for designing better machine learning models and verification of physical laws in machine learning as well.

\appendix
\section{Experiment details}\label{app-a}
\subsection{Forward process}
In this section, we give an example of calculating the entropy change in one-dimensional SDE.
The one-dimensional SDE reads
\begin{equation}
    {\rm{d}}{X_t} =  - {X_t}\;{\rm{d}}t + \sqrt 2 \;{\rm{d}}W,
\end{equation}
with a given initial distribution:
\begin{equation}
    p(X_0) = \frac{1}{2} \cdot \frac{1}{{\sqrt {2\pi } }}{e^{ - \frac{{{{\left( {{X_0} - \mu} \right)}^2}}}{2}}} + \frac{1}{2} \cdot \frac{1}{{\sqrt {2\pi } }}{e^{ - \frac{{{{\left( {{X_0} + \mu} \right)}^2}}}{2}}}.
\end{equation}

We consider the Ito convention $\lambda = 0$. Thus, we have the following discretized equation:
\begin{equation}\label{dsde}
    {X_{t + dt}} = {X_t} - {X_t}\;{\rm{d}}t + \sqrt 2 \;\sqrt {dt} \eta ,
\end{equation}
where $\eta  \sim \N(0,1)$ (i.i.d. in time), $T = 10$, and $dt = 0.01$. We sample from the initial distribution and run Eq.~\eqref{dsde} for a duration of $1000$ steps to obtain an ensemble of stochastic trajectories ${\X}\in\mathbb{R}^{1\,000\times M}$, where $M$ is the number of total trajectories. We define ${X_\tau}(n)$ is the $n$-th element of the column vector for $\tau$-th trajectory.

We calculate the entropy change of the system as
\begin{equation}
   \Delta S = \ln \left[ {\frac{{\exp ( - \frac{{{{({X_\tau }(1) - \mu)}^2}}}{2}) + \exp ( - \frac{{{{({X_\tau }(1) + \mu)}^2}}}{2})}}{{\exp ( - \frac{{{{({X_\tau }(1001) - {\mu e^{ - T}})}^2}}}{2}) + \exp ( - \frac{{{{({X_\tau }(1001) + {\mu e^{ - T}})}^2}}}{2})}}} \right].
\end{equation}
The environment entropy change is calculated as follows,
\begin{equation}
\begin{aligned}
    \Delta {S_E} &= \int_0^T d t \mathop  \odot \limits^{\frac{1}{2}}   [ - \dot XX]\\ &= \sum\limits_{n = 1}^{1000}  -  \left[\frac{{{X_\tau }(n + 1) - {X_\tau }(n)}}{{dt}}\cdot\frac{{{X_\tau }(n + 1) + {X_\tau }(n)}}{2}\right]dt.
\end{aligned}
\end{equation}
The total entropy change can be calculated using
\begin{equation}
    \Delta {S_{tot}} = \Delta {S}_E + \Delta S.
\end{equation}
We sampled $10\,000$ trajectories of the discrete Langevin dynamics. From these trajectories, we compute the statistics of all three entropy quantities and verify the integral fluctuation theorem.  When verifying the convergence of $\lrbka{e^{-\Delta S_{tot}}}$, we change the size of the ensemble.

\subsection{Backward process}
We also use a one-dimensional example to calculate the entropy quantities. 
First, the one-dimensional backward SDE reads
\begin{equation}
   d{{\tilde X}_s} = \left[ {2\tanh \left( {\mu _{T - s}{{\tilde X}_s}} \right)\mu _{T - s} - {{\tilde X}_s}} \right]ds + \sqrt 2 dW,
\end{equation}
where the initial distribution is specified by
\begin{equation}
       p(\tilde X_0) = \frac{1}{{2\sqrt {2\pi } }}\exp \Bigl( - \frac{{{{(\tilde X_0 - {\mu e^{ - T}})}^2}}}{2}\Bigr) + \frac{1}{{2\sqrt {2\pi } }}\exp\Bigr ( - \frac{{{{(\tilde X_0 + {\mu e^{ - T}})}^2}}}{2}\Bigl).
\end{equation}
We consider the Ito convention $\lambda = 0$. Then, the discretized SDE reads
\begin{equation}
    {{\tilde X}_{s + ds}} = {{\tilde X}_s} + \left[ {2\mu\tanh \left( {{\mu e^{ - (T - s)}}{{\tilde X}_s}} \right){e^{ - (T - s)}} - {{\tilde X}_s}} \right]ds + \sqrt 2 \sqrt {ds} \eta ,
\end{equation}
where $\eta  \sim \N(0,1)$ independently for every time step, $T = 10$, and $ds = 0.01$. We start from $\tilde{X}_0$, and run the reverse dynamics for a total of $1\,000$ steps and get a trajectory vector ${{\tilde \X}_\tau }$. We define ${{\tilde X}_\tau }(n)$ as the $n$-th step of the trajectory vector.

The entropy change of the system can be estimated as
\begin{align}
\Delta S = \ln \left[ {\frac{{\exp ( - \frac{{{{({{\tilde X}_\tau }(1) - {\mu e^{ - T}})}^2}}}{2}) + \exp ( - \frac{{{{({{\tilde X}_\tau }(1) + {\mu e^{ - T}})}^2}}}{2})}}{{\exp ( - \frac{{{{({{\tilde X}_\tau }(1001) - \mu)}^2}}}{2}) + \exp ( - \frac{{{{({{\tilde X}_\tau }(1001) + \mu)}^2}}}{2})}}} \right].
\end{align}
We calculate the entropy change of the environment as follows
\begin{equation}
\begin{aligned}
    &\Delta {S_E}  = \int_0^T ds\mathop  \odot \limits^{\frac{1}{2}}    \left[ \dot{\tilde{X}}_s (2\tanh ({\mu _{T - s}} \tilde{X}_s) \mu_{T - s} - \tilde{X}_s) \right]\\ &  = \sum\limits_{n = 1}^{1000} {\left[\frac{{{{\tilde X}_\tau }(n + 1) - {{\tilde X}_\tau }(n)}}{{ds}}\right]} \\&  \times \left[2\mu\tanh ({\mu e^{ - (T - n \cdot ds)}}\frac{{{{\tilde X}_\tau }(n + 1) + {{\tilde X}_\tau }(n)}}{2}){e^{ - (T - n \cdot ds)}}\right.\\ 
    &\left. - \frac{{{{\tilde X}_\tau }(n + 1) + {{\tilde X}_\tau }(n)}}{2}\right]ds.
\end{aligned}
\end{equation}
 We calculate the total entropy production as follows,
\begin{equation}
    \Delta {S_{tot}} = \Delta {S_E} + \Delta S.
\end{equation}
These three entropy quantities are estimated from an ensemble of $10\,000$ stochastic trajectories.
When verifying the convergence of $\lrbka{e^{-\Delta S_{tot}}}$, we change the size of the ensemble. All codes used in this paper are accessible through our GitHub link~\cite{Yu-2024}.

\section{Speciation time for one dimensional example of arbitrary data mean and variance}\label{app-bb}
\blue{According to the analysis in Sec.~\ref{pofree}, we have the following potential function:
\begin{equation}
U(X_t,t)=-\frac{1}{2}X_t^2-2\ln p(X_t,t),
\end{equation}
where $p(X_t,t)$ can be obtained from a similar convolution to that in Eq.~\eqref{eq5} with the result $p(X_t,t)=\frac{1}{2}\left[\mathcal{N}(\mu_t,1+e^{-2t}(\sigma^2-1))+\mathcal{N}(-\mu_t,1+e^{-2t}(\sigma^2-1))\right]$.

It is then verified that the gradient of the log-likelihood is given by
\begin{equation}
\nabla\ln p(X_t,t)=\frac{{{e^t}\left[ { - {e^t}{X_t} + {\mu}\tanh \left( {\frac{{{e^t}{\mu}{X_t}}}{{ - 1 + {e^{2t}} + {\sigma ^2}}}} \right)} \right]}}{{ - 1 + {e^{2t}} + {\sigma ^2}}}.
\end{equation}
A step-by-step calculation is presented in Appendix~\ref{app-d}. Therefore, the first derivative of the potential is given by
\begin{equation}\label{fdpot}
\frac{\partial U(X_t,t)}{\partial X_t}=-X_t+\frac{2X_t}{1+e^{-2t}(\sigma^2-1)}-\frac{2\mu e^{t}\tanh\left(\frac{\mu e^tX_t}{e^{2t}+\sigma^2-1}\right)}{e^{2t}+\sigma^2-1}.
\end{equation}
The second derivative is then given as follows:
\begin{equation}
\frac{\partial^2 U(X_t,t)}{\partial X_t^2}=-1+\frac{2}{1+e^{-2t}(\sigma^2-1)}-2\left(\frac{\mu e^t}{e^{2t}+\sigma^2-1}\right)^2\left[1-\tanh^2\biggl(\frac{\mu e^tX_t}{e^{2t}+\sigma^2-1}\biggr)\right].
\end{equation}

Finally, we check that there are four solutions to $\frac{\partial^2 U(X_t,t)}{\partial X_t^2}\vert_{X_t=0}=0$. Only one of them gives a real value of time, i.e., $t_S=\frac{1}{2}\ln\left(\mu^2+\sqrt{\sigma^4-2\sigma^2+\mu^4+1}\right)$.  It is surprising that the symmetry breaking can become unstable if the sign of the first derivative of the potential at $X_t\to\infty$ changes from negative to positive. This is determined by the following limit:
\begin{equation}
\lim_{X_t\to\infty}\frac{\partial U(X_t,t)}{\partial X_t}=-\frac{2e^t\mu}{C(\sigma,t)}+\lim_{X_t\to\infty}\frac{2X_t}{e^{-2t}C(\sigma,t)}-X_t,
\end{equation}
where we have used Eq.~\eqref{fdpot}, and $C(\sigma,t)=e^{2t}+\sigma^2-1$. It is clear that when $C(\sigma,t)<2e^{2t}$, the potential shape will change from the one of double minima to an inverted U shape. At the critical line $C(\sigma,t)=2e^{2t}$, the potential shape is still of the double-minima type, because of its negative slope at $X_t=\infty$ (the exact value is given by $-\mu e^{-t}$). Therefore, we determine the critical time as $t_{USB}=\frac{1}{2}\ln(\sigma^2-1)$, which is independent of $\mu$. Moreover, the transition is of the first order.}

\section{Proof of the same state probability for both forward and reverse dynamics }\label{app-b}
The forward OU process is given by
\begin{equation}
d{\X_t} = f({\X_t},t)dt + \sqrt{2}d\mathbf{W},
\end{equation}
where $d\mathbf{W}$ is the Wiener process. The solution of the corresponding FPE is specified by $p(\X_t,t)$. 
The reverse generative diffusion dynamics is described by
\begin{equation}
    d{\tilde{\X}_s} = \left( 2{\nabla \ln p({\tilde{\X}_s},T - s) - f({\tilde{\X}_s},T - s)} \right)ds + \sqrt{2}d\mathbf{W},
\end{equation}
where $s$ is in an increasing order.
The corresponding FPE is given below:
\begin{equation}
\begin{aligned}
\frac{{\partial P({\tilde{\X}_s},s)}}{{\partial s}}& =  - \nabla  \cdot \left[ {\left( {2\nabla \ln p({\tilde{\X}_s},T - s) - f({\tilde{\X}_s},T - s)} \right)P({\tilde{\X}_s},s)} \right]\\& + \sum\limits_{i = 1}^d {\frac{{{\partial ^2}P({\tilde{\X}_s},s)}}{{\partial {\tilde{X}_i}\partial {\tilde{X}_i}}}}.
\end{aligned}
\end{equation}
We then replace $P(\tilde{\X}_s,s)$ by $p(\tilde{\X}_s,T-s)$, and keep others unchanged, obtaining the following result:
\begin{equation}\label{equp}
\begin{aligned}
    &\frac{{\partial p({\tilde{\X}_s},T - s)}}{{\partial s}} =  - \nabla  \cdot \left[ {\left( {2\nabla \ln p({\tilde{\X}_s},T - s) - f({\tilde{\X}_s},T - s)} \right)p({\tilde{\X}_s},T - s)} \right]\\& + \sum\limits_{i = 1}^d {\frac{{{\partial ^2}p({\tilde{\X}_s},T - s)}}{{\partial {\tilde{X}_i}\partial {\tilde{X}_i}}}} \\
   & = \nabla  \cdot [f({\tilde{\X}_s},T - s)p({\tilde{\X}_s},T - s)] - \sum\limits_{i = 1}^d {\frac{{{\partial ^2}p({\tilde{\X}_s},T - s)}}{{\partial {\tilde{X}_i}\partial {\tilde{X}_i}}}} \\& \Rightarrow \frac{{\partial p({\tilde{\X}_s},T - s)}}{{\partial (T - s)}} =  - \nabla  \cdot [f({\tilde{\X}_s},T - s)p({\tilde{\X}_s},T - s)] + \sum\limits_{i = 1}^d {\frac{{{\partial ^2}p({\tilde{\X}_s},T - s)}}{{\partial {\tilde{X}_i}\partial {\tilde{X}_i}}}}. 
\end{aligned}
\end{equation}
Noticing that $\tilde{\X}_s=\X(T-t)$ where $t$ is the forward time, we can conclude that the last equation in Eq.~\eqref{equp} is exactly the same with the forward FPE. Therefore, both processes bear the same state probability distribution, as intuitively expected.

\section{Probability currents for both forward and reverse dynamics }\label{app-c}
In this section, we prove that the probability currents have the same magnitude but opposite directions for forward and reverse processes.
First, one can write down the forward probability current as follows,
\begin{equation}
{\J } = f({\X_t},t)p({\X_t},t) - \nabla p({\X_t},t).
\end{equation}
The force term in the reverse SDE is given by $f_{rev}= 2{\nabla \ln p({\tilde{\X}_s},T - s) - f({\tilde{\X}_s},T - s)}$, whose corresponding probability current reads
\begin{equation}
\begin{aligned}
      {\J^{*}  }  &= \left( {2\nabla \ln p({\tilde{\X}_s},T - s) - f({\tilde{\X}_s},T - s)} \right)P({\tilde{\X}_s},s) - \nabla P({\tilde{\X}_s},s)\\& = \left( {2\nabla \ln p({\tilde{\X}_s},T - s) - f({\tilde{\X}_s},T - s)} \right)p({\tilde{\X}_s},T - s) - \nabla p({\tilde{\X}_s},T - s)\\& =  - f({\tilde{\X}_s},T - s)p({\tilde{\X}_s},T - s) + \nabla p({\tilde{\X}_s},T - s)\\
      &=-\J,
\end{aligned}
\end{equation}
where we have replaced $P({\tilde{X}_s},s)$ by  $p({\tilde{X}_s},T - s)$ as in Sec.~\ref{app-b}.

\section{Entropy production rate for $\sigma^2\neq1$}\label{app-d}
In the case of $\sigma^2\neq1$, the score function can also be computed in an analytic form. We only derive the formulas for the one-dimensional example. The details are given below.
We first compute the score function.
\begin{equation}
\begin{aligned}
   &\nabla \ln p({X_t},t) = \frac{{ -  {\frac{{{e^{ -  {\frac{{{{( - {e^{ - t}}{\mu} + {X_t})}^2}}}{{2(1 - {e^{ - 2t}} + {e^{ - 2t}}{\sigma ^2})}}} }}( - {e^{ - t}}{\mu} + {X_t})}}{{1 - {e^{ - 2t}} + {e^{ - 2t}}{\sigma ^2}}}}  - {\frac{{{e^{ -  {\frac{{{{({e^{ - t}}{\mu} + {X_t})}^2}}}{{2(1 - {e^{ - 2t}} + {e^{ - 2t}}{\sigma ^2})}}} }}({e^{ - t}}{\mu} + {X_t})}}{{1 - {e^{ - 2t}} + {e^{ - 2t}}{\sigma ^2}}}} }}{{{e^{ -  {\frac{{{{( - {e^{ - t}}{\mu} + {X_t})}^2}}}{{2(1 - {e^{ - 2t}} + {e^{ - 2t}}{\sigma ^2})}}} }} + {e^{ -  {\frac{{{{({e^{ - t}}{\mu} + {X_t})}^2}}}{{2(1 - {e^{ - 2t}} + {e^{ - 2t}}{\sigma ^2})}}} }}}}\\
   &=\frac{1}{{1 - {e^{ - 2t}} + {e^{ - 2t}}{\sigma ^2}}}\frac{{ -  {{e^{ -  {\frac{( - {e^{ - t}}{\mu} + {X_t})^2}{{2(1 - {e^{ - 2t}} + {e^{ - 2t}}{\sigma ^2})}}} }}( - {e^{ - t}}{\mu} + {X_t})}  -  {{e^{ -  {\frac{{{{({e^{ - t}}{\mu} + {X_t})}^2}}}{{2(1 - {e^{ - 2t}} + {e^{ - 2t}}{\sigma ^2})}}} }}({e^{ - t}}{\mu} + {X_t})} }}{{{e^{ -  {\frac{{{{( - {e^{ - t}}{\mu} + {X_t})}^2}}}{{2(1 - {e^{ - 2t}} + {e^{ - 2t}}{\sigma ^2})}}} }} + {e^{ -  {\frac{{{{({e^{ - t}}{\mu} + {X_t})}^2}}}{{2(1 - {e^{ - 2t}} + {e^{ - 2t}}{\sigma ^2})}}} }} }}\\
   &=\frac{1}{{1 - {e^{ - 2t}} + {e^{ - 2t}}{\sigma ^2}}}\frac{{ -  {{e^{  {\frac{{  {e^{ - t}}{\mu}{X_t}}}{{1 - {e^{ - 2t}} + {e^{ - 2t}}{\sigma ^2}}}} }}( - {e^{ - t}}{\mu} + {X_t})}  -  {{e^{ -  {\frac{{{e^{ - t}}{\mu}{X_t}}}{{1 - {e^{ - 2t}} + {e^{ - 2t}}{\sigma ^2}}}} }}({e^{ - t}}{\mu} + {X_t})} }}{{{e^{  {\frac{{  {e^{ - t}}{\mu}{X_t}}}{{1 - {e^{ - 2t}} + {e^{ - 2t}}{\sigma ^2}}}} }} + {e^{ -  {\frac{{{e^{ - t}}{\mu}{X_t}}}{{1 - {e^{ - 2t}} + {e^{ - 2t}}{\sigma ^2}}}} }}}}\\
      & = \frac{1}{{1 - {e^{ - 2t}} + {e^{ - 2t}}{\sigma ^2}}}\left[ {\tanh (\frac{{{e^{ - t}}{\mu}{X_t}}}{{1 - {e^{ - 2t}} + {e^{ - 2t}}{\sigma ^2}}})({e^{ - t}}{\mu}) - {X_t}} \right]\\
   &=\frac{{{e^t}\left[ { - {e^t}{X_t} + {\mu}\tanh \left( {\frac{{{e^t}{\mu}{X_t}}}{{ - 1 + {e^{2t}} + {\sigma ^2}}}} \right)} \right]}}{{ - 1 + {e^{2t}} + {\sigma ^2}}}.\label{ssigma}
\end{aligned}
\end{equation}
We then estimate the entropy flux for the forward process as follows,
\begin{equation}
\begin{aligned}
 \phi& = \int {\left( {{X_t}p({X_t},t) + \frac{{\partial p({X_t},t)}}{{\partial {X_t}}}} \right)} {X_t}d{X_t} \\& = \int {{X_t}^2p({X_t},t)} d{X_t} + \int {{X_t}} \frac{{\partial p({X_t},t)}}{{\partial {X_t}}}d{X_t}\\
& =\left\langle X_t^2\right\rangle_p-1.
\end{aligned}
\end{equation}
$p(X_t,t)$ is obtained from Eq.~\eqref{pxt} by setting $d=1$.
The entropy production rates are equal in both forward and reverse generative processes. They can be estimated below:
\begin{equation}
\begin{aligned}
&\pi  = {\pi ^ * } =\int {\frac{{{{\left( {{X_t}p({X_t},t) + \frac{{\partial p({X_t},t)}}{{\partial {X_t}}}} \right)}^2}}}{{p({X_t},t)}}} d{X_t} \\
&  = \int {X_t^2p({X_t},t)d{X_t}}  + \int {2{X_t}\frac{{\partial p({X_t},t)}}{{\partial {X_t}}}d{X_t}}  + \int {\frac{1}{{p({X_t},t)}}{{\left( {\frac{{\partial p({X_t},t)}}{{\partial {X_t}}}} \right)}^2}d{X_t}}   \\
&  = {\left\langle {X_t^2} \right\rangle _p} - 2 - {\left\langle {\frac{{{\partial ^2}\ln p({X_t},t)}}{{\partial {X_t}^2}}} \right\rangle _p},
\end{aligned}
\end{equation}
where
${\left\langle {\frac{{{\partial ^2}\ln p({X_t},t)}}{{\partial {X_t}^2}}} \right\rangle _p} = \int {\frac{{{\partial ^2}\ln p({X_t},t)}}{{\partial {X_t}^2}}p({X_t},t)d{X_t}} $.

The entropy flux for the reverse generative process can be estimated as follows,
\begin{equation}
\begin{aligned}
{\phi ^ * } &= \int {\left( {{X_t}p({X_t},t) + \frac{{\partial p({X_t},t)}}{{\partial {X_t}}}} \right)} {f^ * }({X_t},t)d{X_t}\\& = {\left\langle {{f^ * }({X_t},t){X_t}} \right\rangle _p} + \int {{f^ * }({X_t},t)\frac{{\partial p({X_t},t)}}{{\partial {X_t}}}} d{X_t}\\& = {\left\langle {{f^ * }({X_t},t){X_t}} \right\rangle _p} - {\left\langle {\frac{{\partial {f^ * }({X_t},t)}}{{\partial {X_t}}}} \right\rangle _p}\\& = {\left\langle {2\frac{{\partial \ln p({X_t},t)}}{{\partial {X_t}}}{X_t}} \right\rangle _p} + {\left\langle {{X_t}^2} \right\rangle _p} - 2{\left\langle {\frac{{{\partial ^2}\ln p({X_t},t)}}{{\partial {X_t}^2}}} \right\rangle _p} - 1\\& =  - 3 + {\left\langle {{X_t}^2} \right\rangle _p} - 2{\left\langle {\frac{{{\partial ^2}\ln p({X_t},t)}}{{\partial {X_t}^2}}} \right\rangle _p}.
\end{aligned}
\end{equation}
where $*$ indicates the reverse process, and ${f^ * }({X_t},t) = 2\frac{{\partial \ln p({X_t},t)}}{{\partial {X_t}}} + {X_t}$ for the reverse process.

The system entropy rate for the forward OU process can now be written as follows:
\begin{equation}
\begin{aligned}
    \dot S& = \pi  - \phi 
    \\& =  - 1 - {\left\langle {\frac{{{\partial ^2}\ln p({X_t},t)}}{{\partial {X_t}^2}}} \right\rangle _p}.
\end{aligned}
\end{equation}
The same rate for the reverse process is given as follows:
\begin{equation}
\begin{aligned}
    {{\dot S}^ * } &= {\pi ^ * } - {\phi ^ * }
    \\& =    1 + {\left\langle {\frac{{{\partial ^2}\ln p({X_t},t)}}{{\partial {X_t}^2}}} \right\rangle _p}.
\end{aligned}
\end{equation}
We conclude that $\dot{S}$ and $\dot{S}^*$ are anti-symmetric quantities.

To explicitly compute the above physical quantities, we need to compute the following expectations.
The first one is the second moment of $X_t$, which reads
\begin{equation}
    {\left\langle {X_t^2} \right\rangle _p} = {\mu ^2}{e^{ - 2t}} + 1 - {e^{ - 2t}} + {\sigma ^2}{e^{ - 2t}}.
\end{equation}
The second one $\left\langle {\frac{{{\partial ^2}\ln p({X_t},t)}}{{\partial {X_t}^2}}} \right\rangle _p$ is derived below using 
the score function in Eq.~\eqref{ssigma}.
\begin{equation}
\begin{aligned}
    &{\left\langle {\frac{{{\partial ^2}\ln p({X_t},t)}}{{\partial {X_t}^2}}} \right\rangle _p} = \frac{1}{{1 - {e^{ - 2t}} + {e^{ - 2t}}{\sigma ^2}}}{\left\langle {\frac{{\partial \left[ {\tanh (\frac{{{e^{ - t}}\mu {X_t}}}{{1 - {e^{ - 2t}} + {e^{ - 2t}}{\sigma ^2}}})({e^{ - t}}\mu ) - {X_t}} \right]}}{{\partial {X_t}}}} \right\rangle _p}\\& = \frac{1}{{1 - {e^{ - 2t}} + {e^{ - 2t}}{\sigma ^2}}}{\left\langle {\left[1 - {{\tanh }^2}\left(\frac{{{e^{ - t}}\mu {X_t}}}{{1 - {e^{ - 2t}} + {e^{ - 2t}}{\sigma ^2}}}\right)\right]\frac{{{e^{ - 2t}}{\mu ^2}}}{{1 - {e^{ - 2t}} + {e^{ - 2t}}{\sigma ^2}}} - 1} \right\rangle _p}\\&  = \frac{{{e^{ - 2t}}{\mu ^2}}}{{{{\left( {1 - {e^{ - 2t}} + {e^{ - 2t}}{\sigma ^2}} \right)}^2}}} - \frac{{{e^{ - 2t}}{\mu ^2}{{\left\langle {{{\tanh }^2}(\frac{{{e^{ - t}}\mu {X_t}}}{{1 - {e^{ - 2t}} + {e^{ - 2t}}{\sigma ^2}}})} \right\rangle }_p}}}{{{{(1 - {e^{ - 2t}} + {e^{ - 2t}}{\sigma ^2})}^2}}} - \frac{1}{{1 - {e^{ - 2t}} + {e^{ - 2t}}{\sigma ^2}}}.
\end{aligned}
\end{equation}
Inserting these two expression into the entropy flux and production rate, we can get  the final general results for $\sigma^2\neq1$.
\begin{equation}
\begin{aligned}
 \phi  &= {\mu ^2}{e^{ - 2t}} - {e^{ - 2t}} + {\sigma ^2}{e^{ - 2t}},\\
 \pi &= {\pi ^ * }  = {\mu ^2}{e^{ - 2t}} - {e^{ - 2t}} + {\sigma ^2}{e^{ - 2t}} - 1 - \frac{{{e^{ - 2t}}{\mu ^2}}}{{{{\left( {1 - {e^{ - 2t}} + {e^{ - 2t}}{\sigma ^2}} \right)}^2}}}\\&  + \frac{{{e^{ - 2t}}{\mu ^2}{{\left\langle {{{\tanh }^2}(\frac{{{e^{ - t}}\mu {X_t}}}{{1 - {e^{ - 2t}} + {e^{ - 2t}}{\sigma ^2}}})} \right\rangle }_p}}}{{{{(1 - {e^{ - 2t}} + {e^{ - 2t}}{\sigma ^2})}^2}}} + \frac{1}{{1 - {e^{ - 2t}} + {e^{ - 2t}}{\sigma ^2}}},\\
   {\phi ^ * } & = {\mu ^2}{e^{ - 2t}} - {e^{ - 2t}} + {\sigma ^2}{e^{ - 2t}} - 2 - 2\frac{{{e^{ - 2t}}{\mu ^2}}}{{{{\left( {1 - {e^{ - 2t}} + {e^{ - 2t}}{\sigma ^2}} \right)}^2}}}\\& + 2\frac{{{e^{ - 2t}}{\mu ^2}{{\left\langle {{{\tanh }^2}(\frac{{{e^{ - t}}\mu {X_t}}}{{1 - {e^{ - 2t}} + {e^{ - 2t}}{\sigma ^2}}})} \right\rangle }_p}}}{{{{(1 - {e^{ - 2t}} + {e^{ - 2t}}{\sigma ^2})}^2}}} + \frac{2}{{1 - {e^{ - 2t}} + {e^{ - 2t}}{\sigma ^2}}}.
\end{aligned}
\end{equation}
In the case of $\sigma^2=1$, the above formulas reduce to the following simple results:
\begin{equation}
\begin{aligned}
  \phi  & = {\mu ^2}{e^{ - 2t}},\\
   \pi  = {\pi ^ * } & = {e^{ - 2t}}{\mu ^2}{\left\langle {{{\tanh }^2}({e^{ - t}}\mu {X_t})} \right\rangle _p},\\
    {\phi ^ * } &   =  - {\mu ^2}{e^{ - 2t}} + 2{e^{ - 2t}}{\mu ^2}{\left\langle {{{\tanh }^2}({e^{ - t}}\mu {X_t})} \right\rangle _p}.
 \end{aligned}
\end{equation}
It is also interesting to show that $\lrbka{\tanh(\mu_tX_t)\mu_tX_t}=\mu_t^2$ ($\mu_t=\mu e^{-t}$ in one dimension), which holds only for $\sigma^2=1$.

\begin{acknowledgments}
 This research was supported by the National Natural Science Foundation of China for
Grant Number 12122515 (H.H.), and Guangdong Provincial Key Laboratory of Magnetoelectric Physics and Devices (No. 2022B1212010008), 
and Guangdong Basic and Applied Basic Research Foundation (Grant No. 2023B1515040023).  
\end{acknowledgments}



\end{document}